\documentclass[sigconf]{acmart}
\usepackage{arydshln}
\usepackage{comment}
\usepackage{textcomp}
\usepackage{booktabs,arydshln}
\usepackage{makecell}
\usepackage{colortbl,booktabs}
\usepackage{fontawesome5}
\usepackage{array} 
\usepackage{siunitx}
\usepackage{wrapfig,lipsum,booktabs} 
\usepackage{tikz}
\usepackage{adjustbox}
\usepackage{array} 
\usepackage{enumitem}
\usepackage{dblfloatfix} 
\usepackage{soul} 
\def\mybar#1{%%
  {\color{gray}\rule{#1cm}{6pt}}} 

\makeatletter
\def\adl@drawiv#1#2#3{%
        \hskip.5\tabcolsep
        \xleaders#3{#2.5\@tempdimb #1{1}#2.5\@tempdimb}%
                 #2\z@ plus1fil minus1fil\relax 
        \hskip.5\tabcolsep}
\newcommand{\cdashlinelr}[1]{%
  \noalign{\vskip\aboverulesep
          \global\let\@dashdrawstore\adl@draw
          \global\let\adl@draw\adl@drawiv}
  \cdashline{#1}
  \noalign{\global\let\adl@draw\@dashdrawstore
          \vskip\belowrulesep}}
\makeatother

\newif\ifhidecomments
\hidecommentsfalse
\newcommand{\takeaway}[1]{
\noindent\rule{\linewidth}{0.1pt}
\par\nobreak\noindent\textbf{Takeaways:}
#1
\vspace{-2mm}
\par\nobreak\noindent
\rule{\linewidth}{0.1pt}
}

\AtBeginDocument{%
  \providecommand\BibTeX{{%
    \normalfont B\kern-0.5em{\scshape i\kern-0.25em b}\kern-0.8em\TeX}}}

\copyrightyear{2025}
\acmYear{2025}
\setcopyright{cc}
\setcctype{by}
\acmConference[CHI '25]{CHI Conference on Human Factors in Computing Systems}{April 26-May 1, 2025}{Yokohama, Japan}
\acmBooktitle{CHI Conference on Human Factors in Computing Systems (CHI '25), April 26-May 1, 2025, Yokohama, Japan}
\acmISBN{979-8-4007-1394-1/25/04}
\acmDOI{10.1145/3706598.3714286}
\begin{document}

\title{Understanding Attitudes and Trust of Generative AI Chatbots for Social Anxiety Support}

\author{Yimeng Wang}  
\affiliation{
    \institution{William \& Mary}
    \city{Williamsburg}
    \state{VA}
    \country{USA}}
\email{ywang139@wm.edu}
\orcid{0009-0005-0699-4581} 

\author{Yinzhou Wang}  
\affiliation{
    \institution{William \& Mary}
    \city{Williamsburg}
    \state{VA}
    \country{USA}}
\email{ywang143@wm.edu}
\orcid{0009-0009-6355-2551} 

\author{Kelly Crace}  
\affiliation{
    \institution{University of Virginia}
    \city{Charlottesville}
    \state{VA}
    \country{USA}}
\email{kelly.crace@virginia.edu}
\orcid{0009-0004-9097-4059} 

\author{Yixuan Zhang}  
\affiliation{
    \institution{William \& Mary}
    \city{Williamsburg}
    \state{VA}
    \country{USA}} 
\email{yzhang104@wm.edu}
\orcid{0000-0002-7412-4669}

\renewcommand{\shortauthors}{Wang et al.}

\begin{abstract} %150 words max
Social anxiety (SA) has become increasingly prevalent. Traditional coping strategies often face accessibility challenges. Generative AI (GenAI), known for their knowledgeable and conversational capabilities, are emerging as alternative tools for mental well-being. With the increased integration of GenAI, it is important to examine individuals' attitudes and trust in GenAI chatbots' support for SA. Through a mixed-method approach that involved surveys ($n = 159$) and interviews ($n = 17$), we found that individuals with severe symptoms tended to trust and embrace GenAI chatbots more readily, valuing their non-judgmental support and perceived emotional comprehension. However, those with milder symptoms prioritized technical reliability. We identified factors influencing trust, such as GenAI chatbots' ability to generate empathetic responses and its context-sensitive limitations, which were particularly important among individuals with SA. We also discuss the design implications and use of GenAI chatbots in fostering cognitive and emotional trust, with practical and design considerations.
\end{abstract} 





\maketitle
\section{Introduction}
\label{sec:intro}

Many people experience feelings of social anxiety (SA) from time to time, and some people experience such feelings more frequently than others~\cite{NIMH2022_SocialAnxiety, leary1995social}. SA, characterized by an intense fear of social interactions and a heightened sense of self-consciousness, has become increasingly prevalent in recent years~\cite{Goodwin2020}, exacerbated by the COVID-19 pandemic~\cite{Ranta2023}. The resulting prolonged social isolation and disruptions to normal socializing practices have heightened anxieties and made reintegration into social settings more challenging for many~\cite{kindred2023influence, Ranta2023}. Traditional mental health approaches face various challenges, such as a shortage of mental health professionals, healthcare constraints, and limited accessibility~\cite{Ebert2018}. An increasing number of people are turning to AI chatbots to manage SA~\cite{cbsnews2023_mentalhealth}. Many appreciate that these digital tools offer a judgment-free space for enhancing social confidence, particularly appealing to those with less developed social confidence~\cite{Ta2020}. With the rise of Generative AI (GenAI) and Large Language Models (LLMs), more individuals have shown interest in using GenAI chatbots~\footnote{GenAI chatbots in our context refers to conversational agents powered by generative artificial intelligence technologies, which include general-purpose systems like ChatGPT and specialized mental health applications such as Woebot.} to navigate their SA challenges~\cite{reddit2023_socialanxiety}. While some users report positive experiences and express trust in these technologies, others remain skeptical, questioning their effectiveness and safety~\cite{Blease2023, Choudhury2023}. Given that GenAI is becoming more integrated into mental health support systems, it is important to explore how people perceive and trust these tools, as their perceptions could significantly influence future adoption and design tailored to specific needs.

Despite increasing research on GenAI tools in mental health, most studies have focused on enhancing the trustworthiness of AI models by improving their reliability and accuracy to foster user trust~\cite{Bedu2021, vonEschenbach2021, Glikson2020}. However, such research often overlooks the nuanced nature of user trust, which is heavily influenced by personal experiences and the specific context in which the technology is used. In other words, previous studies have primarily sought to address ``what makes the model trustworthy'' but fail to ask, \emph{``in whose voice is this trustworthiness defined?''} While much research focuses on building trust through model evaluation and technical transparency, the reality is that even a technologically sound system does not automatically translate to user trust~\cite{Chen2023}. 

Additionally, individual differences also significantly heavily influence the extent and manner in which people trust the trustee (i.e., the object of trust, such as technology and interpersonal relationships)~\cite{Asan2020, Sutherland2020, Naber2018, Hancock2023}. For example, research in digital contexts has found that people's prior experiences and attitudes significantly shape their trust in social media~\cite{Zhang2024}, suggesting that trust formation can be heavily influenced by personal and contextual factors. Similarly, prior work in healthcare has shown that patients tend to trust their doctors more when dealing with complex diseases, which indicates that trust reflects their need for reassurance and confidence in their healthcare providers' capabilities~\cite{skirbekk2011mandates}. In short, existing work has highlighted the contextual nature of trust formation and its' situatedness~\cite{Kaplan2021}. And yet, no empirical research has explicitly explored these dynamics of trust in GenAI chatbots within the context of SA. Examining the dynamics of trust in GenAI chatbots for dealing with SA is crucial for understanding to what extent these tools are perceived as trustworthy, feasible, and valuable, as well as allows us to assess the potential for these emerging GenAI chatbots to meet their needs.

In this work, we aim to examine the following research questions (RQs): 
\textbf{RQ1.} What are peoples' perceptions, trust, and experiences with GenAI chatbots in dealing with SA in everyday life, and what are their perspectives on future use?  
\textbf{RQ2.} What factors may influence individuals' perceptions, willingness, or reluctance to use GenAI chatbots to deal with SA? and 
\textbf{RQ3.} How do personal differences, such as variations in the severity of SA symptoms and prior engagements interacting with GenAI, shape perceptions of trust and willingness to use GenAI chatbots for support? 

To answer these questions, we conducted a mixed-methods study, including a survey study ($n=159$) and a follow-up semi-structured interview study ($n=17$). Our quantitative results show 1) a strong correlation between trust in GenAI chatbots and their willingness to use it for SA support, 2) individuals experiencing more severe SA symptoms displayed a significantly higher willingness to use GenAI chatbots for coping with their SA challenges, and 3) extensive and frequent prior experience with GenAI chatbots significantly increased their willingness to use these tools for SA support. Our qualitative analysis further reveals the factors influencing trust and willingness to use GenAI chatbots vary with the severity of SA symptoms and past experiences working with human psychotherapists and GenAI chatbots, reflecting different trust dynamics. 
Specifically, we observed that individuals experiencing more severe SA symptoms have shown a higher willingness to use GenAI chatbots for support, primarily valuing \emph{emotional trust}---GenAI chatbots' non-judgmental support and their capacity to simulate empathy. Such emotional trust is crucial for those who rely on GenAI chatbots for emotional engagement and support, transcending mere technical functionality. However, participants with milder SA symptoms in our study prioritized cognitive trust, focusing on the technical reliability and accuracy of the AI models, and competency and the soundness of their operations, as well as the importance of power dynamics and user control in interactions with human-LLM systems, which were considered particularly crucial in forming trust perceptions and determining the willingness to engage with GenAI chatbots for dealing with SA. We also found that some participants with extensive experience with human psychotherapists found GenAI chatbots to be preferable to inadequate human psychotherapists, highlighting a unique niche where GenAI can serve as a consistent, albeit basic, support system.

In this work, our contributions include: 
\textbf{1) } we conducted a mixed-methods study with 159 survey respondents and 17 interview participants to gain an in-depth understanding of individuals' attitudes of, trust in, and intentions to use GenAI chatbots for SA support;
\textbf{2) } we present a nuanced analysis of trust dynamics in GenAI chatbots in terms of how trust in GenAI chatbots is characterized within the context of coping with SA, as well as unpacking the interplay between symptom severity and past experiences with GenAI; and 
\textbf{3) } we provide insights and design implications for future design leveraging GenAI, such as supporting emotional trust building and considering context-specific trust-building strategies.

%----------------------------------------------------------------------------------------------------
\section{Background and Related Work}

\subsection{General Background on Social Anxiety}

SA presents a paradox where individuals fear social interactions yet often deeply desire them~\cite{Goodman2021}, which indicates the inherently social nature of the issue: the avoidance of others is not driven by a lack of interest, but by an overwhelming fear of judgment or rejection that overshadows the need for connection~\cite{quora2024_socialanxiety}. More than just a fear of interaction, SA stems from a fear of failure in these interactions---a fear of being seen as inadequate, awkward, or flawed in the eyes of others. Such internal struggle leaves individuals caught between their longing for social bonds and the intense fear of embarrassment or rejection. Avoidance arises not from disinterest, but as a defensive response to perceived threats in social situations. Although individuals may look for social engagement, their anxiety could distort simple interactions, turning them into sources of fear. The conflict between wanting social connection and fearing negative outcomes creates a cycle of avoidance, which further deepens isolation and exacerbates SA~\cite{Cruz2023}.

The COVID-19 pandemic has intensified the challenges associated with SA. The widespread social isolation and disruption to everyday life left many people struggling to reintegrate into social settings post-pandemic~\cite{santomauro2021global}. These prolonged periods of isolation exacerbated feelings of anxiety, particularly for those already prone to SA. With in-person interactions restricted during the pandemic, many individuals turned to virtual mental health interventions as an alternative to traditional psychotherapy. Even after the pandemic, this shift has persisted, as people have grown accustomed to and continue favoring these virtual options~\cite{Kathirvel2020}. As such, there is a need for more accessible, scalable, and flexible therapeutic solutions to address the growing demand for helping individuals regain social confidence and deal with social anxiety~\cite{Lattie2022}.

\subsection{AI \& Generative AI in Mental Health}
\subsubsection{Increasing Production and Use of AI in Mental Health.} 
AI chatbots have expanded access to mental health support and have shown the potential to overcome barriers associated with traditional therapy. Initially, chatbots utilized rule-based and retrieval-based systems to provide structured responses, greatly enhancing the accessibility of mental health resources~\cite{Xiao2020, Wang2019, Ma_2024, AbdAlrazaq2020}. With advancements in GenAI technologies, chatbots have evolved to offer more dynamic, context-aware interactions through advanced natural language processing~\cite{Wu2023}. The evolution might be beneficial for individuals with SA, providing a platform for them to engage in naturalistic conversations that mirror challenging social situations~\cite{Franze2023}. In recent years, there has been a rapid expansion in the development and adoption of GenAI chatbots across both academic research and industry~\cite{mckinsey_ai_2023}. Example GenAI chatbots include OpenAI's ChatGPT~\cite{openai_chatgpt_2024}, Woebot~\cite{darcy_2024, woebothealth_2024}, Replika~\cite{replika_2024}, etc have gained attention in the field of mental health over the past two years. As GenAI chatbots become more accessible, many people find that these tools as promising coping strategies to provide natural, conversational interaction that mirror the real-world social situations they struggle with~\cite{heliyon2023}. 
For example, recent work has shown that people with underdeveloped social skills or those who experience occasional social discomfort may seek out chatbots as a non-judgmental, low-pressure way to interact and practice social skills~\cite{Franze2023}. 

\subsubsection{Ethical Considerations of GenAI in Mental Health}
\label{subsubsec:ethical_GenAI_literature}
With the increasing use of AI and GenAI in mental health support (e.g., screening for mental health issues~\cite{Sezgin2024, Sezgin2023}, LLM-powered psychotherapy~\cite{Nie2024, Zainab2024}, mental education~\cite{Sezgin2024}), several ethical challenges emerge~\cite{chenneville2024more, ning2024generative, Saeidnia2024}.  
These ethical considerations have centered around (1) \textit{accountability} that encompasses governance, legal responsibilities, and liability, ensuring that actions and decisions by AI are traceable and justifiable~\cite{Tavory2024, Carr2020}; (2) \textit{autonomy} that demands respect for human decision-making, emphasizing informed consent and human oversight so that individuals retain control over their mental health treatment~\cite{koutsouleris2022promise, Rubeis2022, Elyoseph2024}; (3) \textit{equity} that seeks to eliminate biases and ensure fairness and justice in AI interactions~\cite{Rubeis2022, Tortora2024, Lee2021, Torous2024, Warrier2023}; (4) \textit{integrity} that relates to the honesty and ethical conduct in mental health research and psychotherapy delivery~\cite{Solaiman2024, Tortora2024}; (5) \textit{non-maleficence} focusing on preventing harm, avoiding misleading information, and ensuring the safety and mental well-being of users~\cite{Sarkar2023, DeFreitas2023}; (6) \textit{privacy} that focuses on handling mental health data and protection of client confidentiality~\cite{Rubeis2022, Lee2021, Torous2024}; (7) \textit{security} that aims to protect sensitive data from unauthorized access and breaches, emphasizing confidentiality and safety~\cite{Inkster2023, Chen2024}; (8) \textit{transparency} that involves reasoning behind AI-driven mental health recommendations be explainable and accessible to clients and practitioners~\cite{Lee2021, Chen2024}; and (9) \textit{trust}, cultivated through consistent reliability and therapeutic value of AI tools within mental health care~\cite{Solaiman2024}.

In addition, recent studies have begun to empirically examine the ethics of GenAI in mental health contexts. For example, Haman at el.~\cite{haman2023behind} and Heston et al.~\cite{heston2023safety} investigated how likely GenAI is to promote or recognize risky behaviors (e.g., encouraging substance use~\cite{haman2023behind} or detecting suicidality~\cite{heston2023safety}). Their findings show that, although the AI did not endorse harmful actions, it responded too slowly when faced with urgent user messages that warranted immediate referral to health services. De Freitas et al.~\cite{de2024chatbots} examined how users perceive AI-generated responses and found that users reacted negatively to messages they deemed harmful. These adverse responses included nonsensical or irrelevant replies that neglected sensitive user inputs, as well as harmful language or suggestions that could cause distress~\cite{de2024chatbots}.

However, practical implementation of ethical principles often remains challenging. For example, a recent article from the New York Times~\cite{nytimes2024characterai} reported a teenager’s suicide possibly linked to interactions with a character-driven GenAI chatbot (Characters.AI~\cite{charactersAI}, a neural language model chatbot that generates human-like text responses and participates in contextual conversation). This news article suggests the serious risks of deploying such GenAI tools without sufficient oversight or understanding of their limitations, as well as raises critical questions about the responsibility of developers and service providers to anticipate and mitigate risks, especially when these tools are marketed for emotionally sensitive contexts. Despite the growing utilization and ethical attention of GenAI in mental health, research specifically focusing on its application to SA remains limited. It is important to avoid the assumption that the attitudes to general mental health are similar across all mental health conditions~\cite{Veinot2018}. Our work aims to address this gap by exploring people's attitudes, perceptions, and intentions of using GenAI chatbots in dealing with SA.

\subsection{Situated Perspective of Trust Dynamics in Generative AI}

\subsubsection{Trust in AI \& GenAI} Trust is a critical factor in the adoption and efficacy of AI systems across various domains~\cite{Lukyanenko2022}. In general, trust in AI refers to users' confidence in AI's reliability, transparency, and safety~\cite{Wang2024}. For users to engage with AI, especially in sensitive contexts such as mental health, it is essential that users perceive the system as trustworthy. Prior work has shown a variety of factors could contribute to trust in AI systems, such as transparency about how the technology operates, the security of personal data, and the perceived effectiveness~\cite{Mller2023}. For example, when users believe that their data is secure and that the AI operates with integrity, they are more likely to rely on it for tasks that impact their well-being~\cite{Mehrotra2024}. As AI evolves into more advanced forms such as GenAI, which can generate nuanced responses in ongoing interactions, the complexity of trust dynamics also grows~\cite{Bhattacharya1998}. The increased complexity arises from the LLMs' ability to adapt and respond in ways that more closely mimic human interaction~\cite{Zhou2023}, potentially deepening the emotional connection between users and the GenAI~\cite{Pelau2021}, which could impact user trust. As such, this paradigm shift requires us to rethink the process of trust-building in GenAI, which requires a deeper exploration of how users perceive these advanced GenAI systems.

\subsubsection{Situatedness of Trust Dynamics.} 
Trust is also highly contextual, varying significantly based on who defines trustworthiness, under what circumstances, and with what consequences~\cite{Veinot2018}. In other words, studying trust requires researchers to unpack the nuances of trust dynamics from a \emph{situated perspective}. When considering the situatedness of trust perceptions, we describe a couple of aspects to examine the situatedness of trust dynamics. 

First, the concept of situatedness emphasizes that trust is not a static attribute inherent to the technology (i.e., in GenAI chatbots in our context), but is influenced by the specific contexts in which the technology is deployed~\cite{Kaplan2021}. While a great body of work has extensively examined the trustworthiness of AI models to build user trust in AI~\cite{Bach2022}, often, the results show that the trustworthiness of AI models does not automatically translate into user trust~\cite{Yildirim2021}. In other words, even a technologically sound model does not automatically ensure user trust, which can vary significantly across different contexts and applications. For example, in financial services, users may distrust AI not due to its functionality but because of concerns over data privacy or decision-making transparency~\cite{Yu2022}. As such, trust is not merely a characteristic of the AI system itself but is dynamically formed and influenced by the user's context, personal experiences, and specific application scenarios. In settings such as mental health support, users may require more than just technological sophistication to place their trust in the system. 

Second, personal factors also significantly shape trust dynamics, such as health conditions and prior experiences with technologies~\cite{Asan2020}. For example, prior research has demonstrated a relationship between trust in AI and willingness to use AI in healthcare settings~\cite{vanderHulst2023}. A recent survey study evaluated healthcare professionals' trust in and intention to use GenAI chatbots in decision-making and found that clinicians who saw AI as a tool to reduce their workload were more likely to trust and incorporate it into clinical practice~\cite{Shamszare2023}. Likewise, other work has shown that trust in AI-recommended medical diagnoses can vary depending on the patient's severity of condition or their technological literacy~\cite{Birkhuer2017}. This prior work has collectively suggested that individuals with different levels of health conditions had different perceptions and interactions with technologies, which also affected their trust and their willingness to use them for support. 

Our study extends the existing literature by focusing on the context of the emerging GenAI innovations in the support for SA among individuals with various levels of symptoms. So far, very little work has closely examined such trust dynamics from a situated perspective in the context of mental health and particularly in SA. Our study seeks to address this research gap. In this work,
we aim to deepen the understanding of trust dynamics, focusing on the perceptions, trust, and willingness to use GenAI chatbots for dealing with SA challenges in their daily life.
Understanding these nuances can guide the development of effective and trustworthy solutions tailored to address the unique challenges of specific mental health needs.

% %----------------------------------------------------------------------------------------------------
\section{Methods}
Upon approval from the Institutional Review Board (IRB) at our institution, we conducted a mixed-method study, including a survey study ($n=159$) and a follow-up interview study ($n=17$) to understand individuals' perceptions, trust, and willingness to use GenAI chatbots for potential SA support. Below, we describe the methods of the survey (\autoref{sec:survey_study}) and interview (\autoref{sec:interview_study}) study.  

\subsection{Survey Study}
\label{sec:survey_study}

\subsubsection{Survey Study Procedure} 
Between March and April 2024, we conducted a survey study. First, we conducted several rounds of pilot studies, such as testing the survey within our research group, where members suggested changes to the survey questions and response options. After making adjustments based on their feedback, we conducted another pilot test with a small group of 20 participants to gather further suggestions.

Our inclusion criteria include participants to be aged 18 or older, who either experience varying levels of SA, from mild discomfort to severe symptoms, as well as individuals interested in the potential of GenAI chatbots for social anxiety support. 
We advertised our study at our institution through multiple methods, such as digital flyers and mailing lists. Additionally, we used a snowball sampling method, encouraging individuals to share the survey invitation with peers who met the study's criteria. To determine the sample size, we conducted a power analysis, considering a moderate effect size (Cohen’s f = 0.3), anticipated group differences, acceptable error rates, and the exploratory nature of the study. Our power analysis shows the estimated sample size was 29 participants per group.
In total, we collected 332 responses, with 159 participants completing the survey and meeting the criteria for consistent responses.

\subsubsection{Survey Measures}
The survey study has focused on respondents' SA experiences, interaction with and trust in GenAI chatbots, and their demographic background (see more details in \autoref{tab:demographic}). We summarize the dimensions of survey measures below (see \autoref{tab:survey_measures}), with detailed questions in \autoref{sec:appendices}. 

\begin{table*}[ht!] 
\small
\centering
  \caption{Summary of Survey Measures.}
  \label{tab:survey_measures}
  \begin{tabular}{p{3.5cm}p{3.5cm}p{8.5cm}}
    \toprule
    \textbf{Category} & \textbf{Measures} & \textbf{Details}  \\ 
    \midrule
    Social Anxiety & Diagnosis & Have you been diagnosed with social anxiety by a healthcare professional?  \\ 
    & Self-identification & Do you consider yourself to have social anxiety? \\ 
    & Social Phobia Inventory (SPIN) & 17-item scale assessing social anxiety severity (0–4 per item, total 0–68).  \\ 
    \midrule
    Coping Strategies & External Support & Have you sought external support for managing social anxiety?  \\ 
    & Barriers to Seeking Help & Select barriers experienced when seeking mental health help.\\ 
    \midrule
    Past Experiences & Duration & How long have you been using GenAI chatbots?  \\ 
    & Frequency & How often have you used GenAI chatbots for SA in the past 6 months?  \\ 
    \midrule
    Willingness to Use & Openness & Would you be open to using a GenAI chatbot for managing social anxiety?  \\ 
    \midrule
    Trust & Competence & I believe GenAI chatbots provide accurate and helpful information.  \\ 
    & Honesty & I trust GenAI chatbots to handle my personal information securely.  \\ 
    & Experience & I would follow recommendations from GenAI chatbots even if unsure.  \\ 
    & Benevolence & I believe GenAI chatbots act in my best interest.  \\ 
    & Reliability & I feel GenAI chatbots understand my needs and respond appropriately. \\ 
    & Expectation & I think GenAI chatbots will become more trustworthy as technology advances. \\ 
    \midrule
    Demographic Background & Age & Numeric data grouped into categories.  \\ 
    & Gender & Self-reported gender identity.  \\ 
    & Education & Highest degree completed.  \\ 
    & Race & Self-identified racial background.  \\ 
    \bottomrule   
  \end{tabular} 
\end{table*}

\textbf{Social Anxiety.}  
Based on discussions with our mental health collaborators, we integrated the following three aspects of measurement of SA, with cross-validated participants' self-reports (i.e., their self-identification with the condition), clinical diagnosis (i.e., participants reported on their formal diagnosis of SA), standardized SPIN scores (i.e., scores on the Social Phobia Inventory (SPIN)~\cite{Connor_Davidson_Churchill_Sherwood_Weisler_Foa_2000}, a widely used diagnostic assessment for SA ) to ensure consistency. To ensure validity, we excluded participants whose responses were inconsistent (e.g., those who reported a history of SA, and self-identified as having SA but scored negligible on the SPIN). We then used the SPIN scale as the variable to define the severity of participants' SA symptoms.

\textbf{SA Coping Strategies.} We collected information on participants' 1) coping strategies for managing SA, by asking about their usage of external support for managing SA (i.e., usage of external support or resources, such as psychotherapy, medication, GenAI chatbots, digital mental health apps, and others); and 2) barriers to seeking help (i.e., cultural limitations, social pressures, financial constraints, feelings of shame, fear of stigma, and more).

\textbf{Past Experiences with GenAI Chatbots.} We examined participants' past experience of interacting with GenAI chatbots by asking about 1) their usage length (i.e., how long participants have been using GenAI chatbots, ranging from those who have never used them to those with more than one year of experience.) and 2) usage frequency (i.e., asking how often participants used GenAI chatbots specifically for SA support in the past six months, with response options indicating varying levels of frequency from never to several times a week).

\textbf{Willingness to Use GenAI Chatbots for SA Support.} We evaluated participants' openness to using GenAI chatbots as tools for SA support by examining their general willingness (i.e., by showing options of \textit{No, I would not be interested; Yes; Possibly, I am open to use}).

\textbf{Trust in GenAI Chatbots for SA Support.} Trust in GenAI chatbots was assessed using six survey items on a five-point Likert scale, with responses ranging from strongly disagree (1) to strongly agree (5). The trust measurement items correspond to the trust dimensions supported by literature~\cite{Glikson2020, Zhang2024, zhang2023we}, collectively assess the key dimensions of trust in GenAI chatbots, including competence, honesty, experience, benevolence, reliability, and future expectations, by gauging participants' confidence in the chatbot's ability to provide accurate information, protect personal data, respond appropriately to needs, and act in their best interest, both now and as technology advances.

\textbf{Demographic Background.} We also asked participants to provide information about their age, gender, race, and education to help us understand the overall characteristics of survey respondents.

\begin{table*}[ht!] 
\footnotesize
\centering
  \caption{Participant Characteristics.}
  \label{tab:demographic}
  \begin{tabular}{p{3.5cm}p{4cm}rrlrrl}
    \toprule
        \textbf{Dimension} & \textbf{Response Options} & \textbf{Survey} & &  & \textbf{Interview} & & \\ 
        & & \textbf{Respondents} &  &  & \textbf{Participants} &  & \\ 
        & & \textbf{(N=159)} & \textbf{\%} & & \textbf{(N=17)} & \textbf{\%} & \\ 
    \midrule
    Age & 18-24 & 118 & 74\% & \mybar{.74} & 13 & 76\% & \mybar{.76} \\
         & 25+   & 41 & 26\% & \mybar{.26} & 4 & 24\% & \mybar{.24} \\ 
         \cdashlinelr{2-8}
         & * Mean = 23, SD = 3.5   \\
    \midrule
    Gender & Man & 76 & 48\% & \mybar{.48} & 8 & 47\% & \mybar{.47} \\  
           & Woman & 81 & 51\% & \mybar{.51} & 9 & 53\% & \mybar{.53} \\
           & Non-binary & 1 & 0.6\% & \mybar{.006} & NA  &  & \\ 
           & Prefer not to answer & 1 & 0.6\% & \mybar{.006} & NA  &  & \\
    \midrule
    Education & High school graduate & 36 & 23\% & \mybar{.23} & 4 & 24\% & \mybar{.24} \\
              & Associate degree & 35 & 22\% & \mybar{.22} & 4 & 24\% & \mybar{.24} \\ 
              & Bachelor's degree & 42 & 27\% & \mybar{.27} & 6 & 35\% & \mybar{.35} \\ 
              & Master's degree & 34 & 21\% & \mybar{.21} & 3 & 18\% & \mybar{.18} \\ 
              & Doctorate or professional degree & 9 & 6\% & \mybar{.06} & NA &  & \\
              & Prefer not to answer & 2 & 1\% & \mybar{.01} & NA  &  & \\ 
    \midrule
    Race & White & 48 & 30\% & \mybar{.3} & 5 & 29\% & \mybar{.29} \\
         & African American or Black & 8 & 5\% & \mybar{.05} & 1 & 6\% & \mybar{.06} \\
         & Asian & 97 & 61\% & \mybar{.61} & 11 & 65\% & \mybar{.65} \\
         & Other & 3 & 2\% & \mybar{.02} & NA  &  & \\
         & Prefer not to answer & 3 & 2\% & \mybar{.02} & NA  &  & \\
    \midrule
    SA diagnosis 
    & Yes & 24 & 15\% & \mybar{.15} & 4 & 24\% & \mybar{.24} \\
    & No & 135 & 85\% & \mybar{.85} & 13 & 76\% & \mybar{.76} \\
    \midrule
    Self-identified SA 
    & Yes & 50 & 31\% & \mybar{.31} & 7 & 41\% & \mybar{.41} \\
    & No & 71 & 45\% & \mybar{.45} & 4 & 24\% & \mybar{.24} \\
    & Not sure & 38 & 24\% & \mybar{.24} & 6 & 35\% & \mybar{.35} \\
    \midrule
    Symptoms of SA 
                  & No social anxiety (0-20) & 65  & 40\% & \mybar{.4} & 3 & 18\% & \mybar{.18} \\
    (SPIN scores) & Mild (21-30)             & 28 & 18\% & \mybar{.18} & 4 & 23\% & \mybar{.23} \\
                  & Moderate (31-40)         & 25 & 16\% & \mybar{.16} & 3 & 18\% & \mybar{.18} \\
                  & Severe (41-50)           & 14 & 9\% & \mybar{.09} & 3 & 18\% & \mybar{.18} \\
                  & Very severe (51-68)      & 27 & 17\% & \mybar{.17} & 4 & 23\% & \mybar{.23} \\ 
    \midrule
    Duration of GenAI chatbot use 
    & Never & 15 & 10\% & \mybar{.1} & NA  &  & \\ 
    & Less than 3 months & 21 & 13\% & \mybar{.13} & 1 & 6\% & \mybar{.06} \\ 
    & 3-6 months & 30 & 19\% & \mybar{.19} & 4 & 24\% & \mybar{.24} \\ 
    & 7-12 months & 34 & 21\% & \mybar{.21} & 7 & 41\% & \mybar{.41} \\ 
    & More than one year & 59 & 37\% & \mybar{.37} & 5 & 29\% & \mybar{.29} \\ 
    \midrule
    Frequency of GenAI chatbot use & Never & 127 & 80\% & \mybar{.8} & 11 & 65\% & \mybar{.65} \\ 
    for SA Management & Rarely (once or twice a month) & 21 & 13\% & \mybar{.13} & 4 & 24\% & \mybar{.24} \\ 
    & Occasionally (a few times a month) & 4 & 2.5\% & \mybar{.025} & 2 & 12\% & \mybar{.12} \\ 
    & Frequently (once or twice a week) & 4 & 2.5\% & \mybar{.025} & NA  &  & \\ 
    & Very frequently (several times a week) & 3 & 2\% & \mybar{.02} & NA   &  & \\ 
  \bottomrule   
\end{tabular} 
\end{table*}

\subsubsection{Survey Data Analysis} 
We conducted our survey data analysis between April and May, 2024. To answer our RQs (detailed in \autoref{sec:intro}), we focus on trust, willingness to use, and the severity of SA symptoms, informed by prior work that demonstrates the relationship between trust in AI and willingness to use it in healthcare settings~\cite{vanderHulst2023}, with willingness being notably contextual, influenced by personal factors such as disease severity or familiarity with technology~\cite{Gaczek2023}. Our study extends this exploration to the context of GenAI chatbots for SA support, investigating how trust and willingness interrelate. 

To gain a general overview, we first performed descriptive statistics to summarize demographic characteristics and key variables, calculating means and standard deviations for continuous data and frequencies with percentages for categorical data. 

To evaluate the reliability and construct validity of our trust scale, we used Cronbach’s $\alpha$ and Pearson's correlation matrix. We then conducted an Exploratory Factor Analysis (EFA) (using R package psych~\cite{psych}) to determine whether the trust scale items loaded onto a single factor, verifying the suitability of our data with the Kaiser-Meyer-Olkin (KMO) measure and Bartlett's test of sphericity. To further explore the relationship between trust in GenAI chatbots and future willingness to use them, we performed Kruskal-Wallis Rank Sum Tests (using R package stats~\cite{stats}) to examine differences between groups. Where significant differences were detected, we conducted post-hoc analyses using Dunnett's test (using the R package FSA~\cite{FSA}) to identify specific group differences. 
We further conducted Linear regression analysis with a Wald test and Spearman's rank correlation to quantify the association between trust and participants' likelihood of adopting GenAI chatbots. Effect sizes were reported to assess the magnitude of the observed associations. For the Kruskal-Wallis test, $\eta^2$ was calculated as the effect size. For Linear regression, $R^2$ was used to indicate the proportion of variance explained, while Spearman’s rank correlation was summarized using $\rho$.

To explore the factors influencing willingness, we conducted an Ordinal Logistic Regression (using R package MASS~\cite{MASS}) to assess the impact of SA severity and GenAI chatbot usage patterns. To dig deeper into the underlying patterns of anxiety severity and GenAI chatbot usage, we applied the Gaussian Mixture Model (GMM) (using R package mclust~\cite{mclust}) to identify distinct participant profiles. This approach allowed us to categorize participants into groups with shared characteristics and to gain insights into user behavior and engagement with GenAI chatbots. Where relevant, we reported effect sizes to indicate the strength of the relationships between variables.

\subsection{Interview Study}
\label{sec:interview_study}
To further explore nuance changes in participants’ attitudes and investigate the trends observed in the survey, we conducted follow-up semi-structured interviews via Zoom and qualitative data analysis between May and August 2024.

\subsubsection{Participant Recruitment}
We first screened all survey respondents who were willing to conduct follow-up interviews, divided them into groups according to the severity of their SA symptoms, the duration they used GenAI chatbots, and the frequency of using GenAI chatbots to cope with SA, and then selected a balanced number of potential participants from each group. We contacted these potential participants ($n = 45$) by email and text and received 25 positive responses, 17 of whom completed the interview study, and 8 participants did not show up or cancel the interview meeting. We also presented the demographic information for interview respondents in \autoref{tab:demographic}.

\subsubsection{Interview Study Procedures}
Before each interview, we reviewed the participants' survey data to identify SA experiences, their trust in GenAI chatbots, and distinctive factors that enhance their willingness to use GenAI chatbots for SA. During the interviews, participants were asked to further explain their experiences with SA, their engagement or consideration of GenAI chatbots for SA support, and their perceptions of and trust in the emerging GenAI chatbots' effectiveness and trustworthiness.  At the end of the study, participants were compensated with a \$20 gift card for their time and effort. The interview sessions were audio-recorded for analysis and lasted approximately 60 minutes. 

\subsubsection{Interview Data Analysis}
All interview sessions were transcribed verbatim. We applied the General Inductive Approach~\cite{thomas2006general} for thematic analysis. First, the lead author carefully reviewed the transcripts to understand the key concepts that surfaced during the interviews, and began the coding process by labeling specific concepts (forming low-level codes, such as aspects that seem to be associated with participants' framing of trust and distrust). These low-level codes were then grouped into broader themes focused on trust and distrust. The entire research team engaged in regular discussions to review, refine, and verify these themes throughout the analysis process to mitigate the risk of bias that could arise from a single researcher’s interpretations.

Through the analysis of survey and interview data, we aim to achieve two forms of data triangulation: convergence and complementarity~\cite{triangulation2014use}. Convergence occurs when there is a significant alignment between the quantitative and qualitative data, enhancing the accuracy and reliability of our findings. Complementarity, on the other hand, allows us to create a more detailed and comprehensive understanding by integrating insights from each method, where the results from one approach inform and enhance the other. The goal of triangulation is to help validate our conclusions and provide a more nuanced perspective to answer our research questions.

%----------------------------------------------------------------------------------------------------
\section{Survey Results}

In this section, we first report the relationship between trust and participants' willingness to use GenAI chatbots (\autoref{sec:trust_willingness}) to answer RQ1. We then examine factors associated with individuals' willingness to use GenAI chatbots for SA support (RQ2) (\autoref{sec:willingness_factor}) and explore participant profiles that shape willingness (RQ3) (\autoref{sec:Cluster_analysis}).

\subsection{Trust and Willingness in Using GenAI Chatbots for SA Support (RQ1)}
\label{sec:trust_willingness}
We explore participants' perceptions and trust with GenAI chatbots in dealing with SA in everyday life, and their perspectives on future use (and to answer RQ1). Below, we first report the reliability and validity of the trust scale used in this study and then present our survey results.

\subsubsection{Reliability and Validity of Trust Scale} 
To evaluate trust in GenAI chatbots for SA support, we adopted six dimensions adapted from established research: competence, honesty, experience, benevolence, reliability, and expectation~\cite{Glikson2020, Zhang2024}. We used standard validation measures, including Cronbach's $\alpha$ and Exploratory Factor Analysis (EFA), to assess the consistency and structural integrity of our trust scale.

\begin{figure}[h!]
\centering
\includegraphics[width=0.7\linewidth]{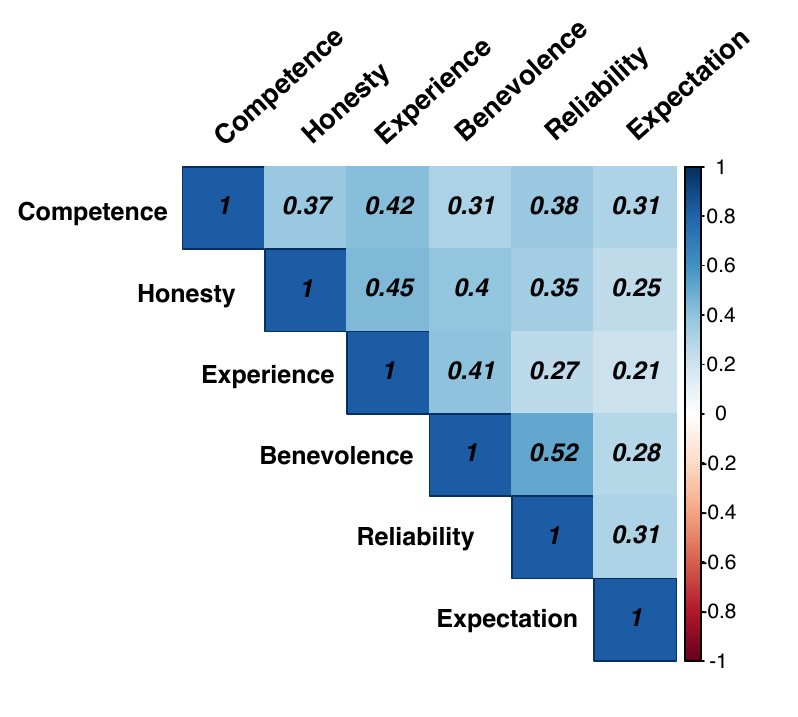}
\caption{Correlation matrix displaying the coefficients between various aspects of trust, including competence, honesty, experience, benevolence, reliability, and expectation) in GenAI chatbots.}
\label{fig:trust_correlation}
\end{figure}

We achieved a Cronbach's $\alpha$ of 0.831, indicating strong internal consistency. Moreover, we created a correlation matrix to explore the interrelationships among trust dimensions. As shown in \autoref{fig:trust_correlation}, the analysis showed positive correlations among all dimensions, indicating that they were interrelated and contributed to the overarching concept of trust in GenAI chatbots. The correlation coefficients ranged from 0.21 to 0.52, indicating distinct contributions from each dimension without significant overlap or redundancy. These results support the multidimensional structure of our trust measure in our study.

To further validate our trust construct, we conducted an EFA to determine whether the six trust dimensions would load onto a single underlying factor. The Kaiser-Meyer-Olkin (KMO) measure of sampling adequacy was 0.79, confirming that our data was suitable for factor analysis. Bartlett's test of sphericity also supported this suitability, and was significant ($\chi^2 =$ 187.262, $df =$ 15, $p <$ 0.001), confirming that the correlations between items were sufficient to proceed with factor analysis. The parallel analysis scree plot (see \autoref{fig:criteria}C in \autoref{sec:appendices}) generated during the EFA suggested that a single factor was appropriate, supported by an eigenvalue greater than one. We used Principal Axis Factor Analysis using the \texttt{fa} function in R, with a \texttt{promax} rotation to account for the potential slight correlations between factors. The analysis confirmed that all six trust aspects loaded strongly onto one factor, with loadings ranging from 0.44 to 0.67 (see \autoref{tab:factor_analysis}). Although the total variance explained (TVE) by the single factor is 0.4, this is considered acceptable given the exploratory nature of our study~\cite{Cutillo2019}. 

\begin{table}[h!]
\centering
\small
\caption{Factor Analysis results on trust measurement.}
\begin{tabular}{p{6cm}>{\raggedleft\arraybackslash}p{1.5cm}}
    \toprule
    Variable & Factor 1 (Trust)\\
    \midrule
    \textbf{Competence:} Ability to provide accurate and helpful information  & 0.60  \\
    \textbf{Honesty:} Handle sensitive or personal information     & 0.62  \\
    \textbf{Experience:} Follow a suggestion or recommendation made by GenAI chatbots  & 0.60   \\
    \textbf{Benevolance:} Working in user's best interest & 0.67  \\
    \textbf{Reliability:} Understand needs and respond appropriately & 0.63   \\
    \textbf{Expectation:} Become more trustworthy as technology improves & 0.44  \\
    \bottomrule
\end{tabular}
\label{tab:factor_analysis}
\end{table}

\subsubsection{Relationship of Trust and Willingness in GenAI Chatbots to Cope with SA} 

In the context of our study, we categorized the willingness to use GenAI chatbots for SA management into three groups: \textit{Willing} (i.e., participants chose \textit{``Yes,''}, \textit{Undecided} (i.e., participants chose \textit{``Possibly, I am open to use''}), and \textit{Unwilling} (i.e.,  \textit{``No, I would not be interested,''}).

\begin{figure}[h!]
\centering
\includegraphics[width=\linewidth]{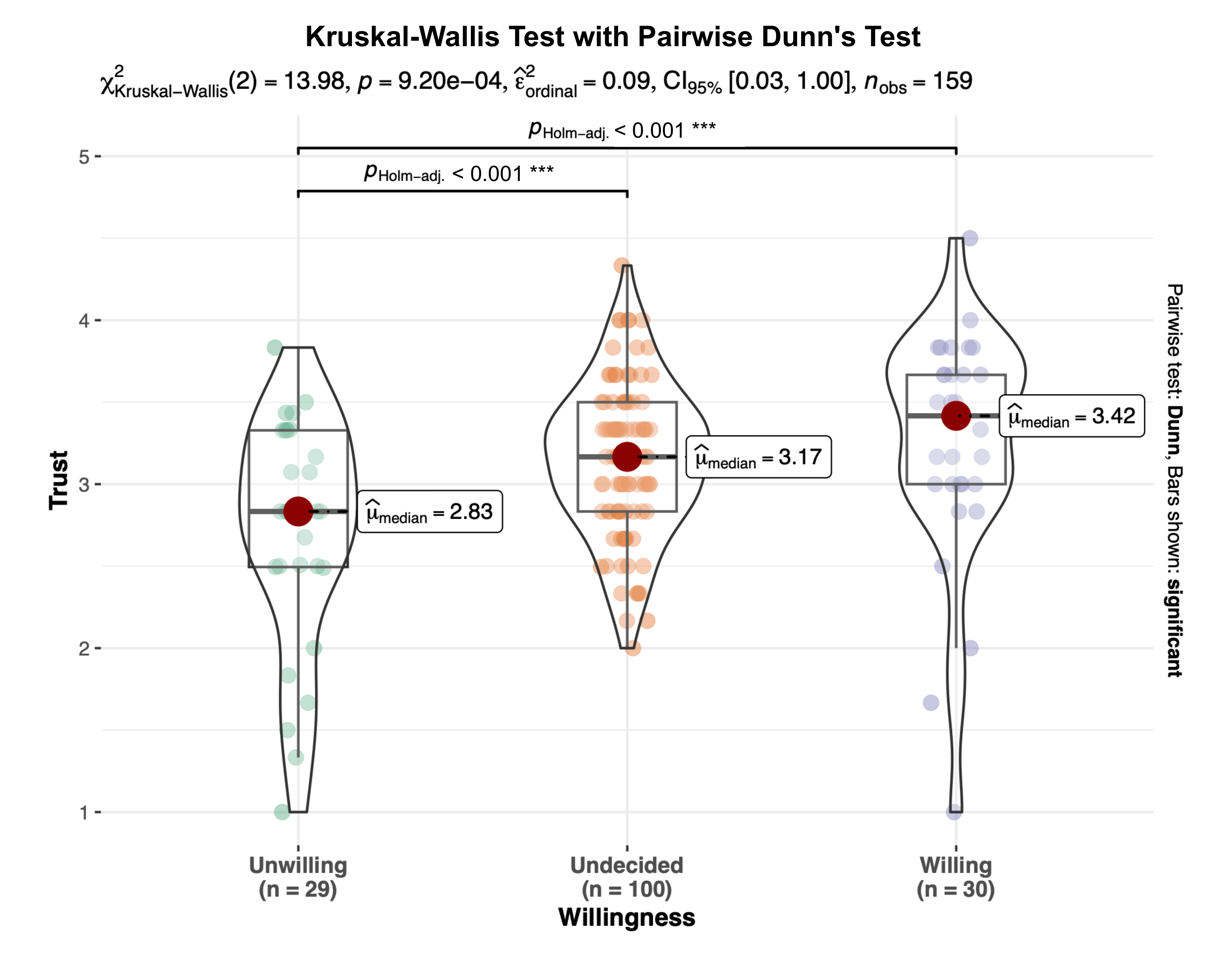}
\caption{The Kruskal-Wallis test assessed differences in trust levels across willingness groups, with a non-parametric pairwise comparison using Dunn's test identifying specific differences between the groups.}
\label{fig:trust_willingness}
\end{figure}

Our analysis revealed a significant variation in trust levels across different willingness groups ($\chi^2 =$ 13.98, $p <$ 0.001). Post-hoc tests (see \autoref{fig:trust_willingness}) further demonstrated that respondents in the \textit{Willing} group (median = 3.42) and the \textit{Undecided} group (median = 3.17) showed significantly higher trust in GenAI chatbots compared to those in the \textit{Unwilling} group, who had a median trust level of 2.83 ($p <$ 0.001). This finding highlights that \textbf{individuals with no intention of using GenAI chatbots in the future show the lowest levels of trust.}

We then performed a linear regression analysis with a Wald test to further explore the relationship between future willingness and trust in GenAI chatbots (see the relationship between the predicted values and the actual values in \autoref{fig:visual}A in \autoref{sec:appendices}). As shown in \autoref{tab:regression_trust}, our results show a \textbf{significant positive association between willingness to use and trust in GenAI chatbots} ($R^2 = 0.13$, which is acceptable in social science~\cite{Ozili2022}). This result aligns with a Spearman's rank correlation, which showed a statistically significant moderate positive association between trust and willingness ($\rho = 0.30$, $p = 0.0003$), indicating that trust tends to increase as willingness to use GenAI chatbots grows. Respondents possibly open to using GenAI chatbots demonstrate a significantly higher level of trust than those without interest, with an increase of 0.649 ($p <$ 0.01). Furthermore, those who expressed a clear willingness to use GenAI chatbots showed an even greater increase in trust, with a coefficient of 0.730 ($p <$ 0.01) compared to the group without interest. These results suggest that \textbf{trust in GenAI chatbots to cope with SA progressively increases as participants' willingness to use shifts from \textit{Unwilling} to \textit{Undecided} and then to \textit{Willing} to use.}

\begin{table}[h!]
    \centering
    \small
    \caption{Linear regression model showing the relationship between respondents' trust in GenAI chatbots and future willingness to use. (Significance level: * $p < 0.05$, ** $p < 0.01$, *** $p < 0.001$)}
    \label{tab:regression_trust}
    \begin{tabular}{p{2.5cm}p{1.4cm}p{1.2cm}p{0.8cm}p{0.9cm}}
    \toprule
     & \multicolumn{4}{c}{\textbf{Trust}} \\
    \cline{2-5}
     & Estimate ($\beta$)& Std. Error & t value & Pr($>$$|$t$|$) \\ 
    \midrule
    \textbf{Willingness to use GenAI chatbots} \\ 
    Intercept & 2.509 & 0.191 & 13.113 & 0.000*** \\
    (\textit{Reference: Unwilling})\\
    Undecided & 0.649 & 0.197 & 3.300 & 0.002** \\
    Willing & 0.730 & 0.234 & 3.118 & 0.002**\\
    \bottomrule
    \end{tabular}
\end{table}

\subsection{Relationship between Severity of SA Symptoms, GenAI Chatbots Usage Patterns, and Willingness to Use GenAI Chatbots for SA Support (RQ2)} 
\label{sec:willingness_factor} 
To answer RQ2, which seeks to explore factors that may influence individuals' perceptions, willingness, or reluctance to use GenAI chatbots in dealing with SA, we examined the relationship between the severity of SA symptoms, GenAI chatbot usage patterns, and people's future willingness to engage with these tools. 

\begin{figure}[h!]
\centering
\includegraphics[width=\linewidth]{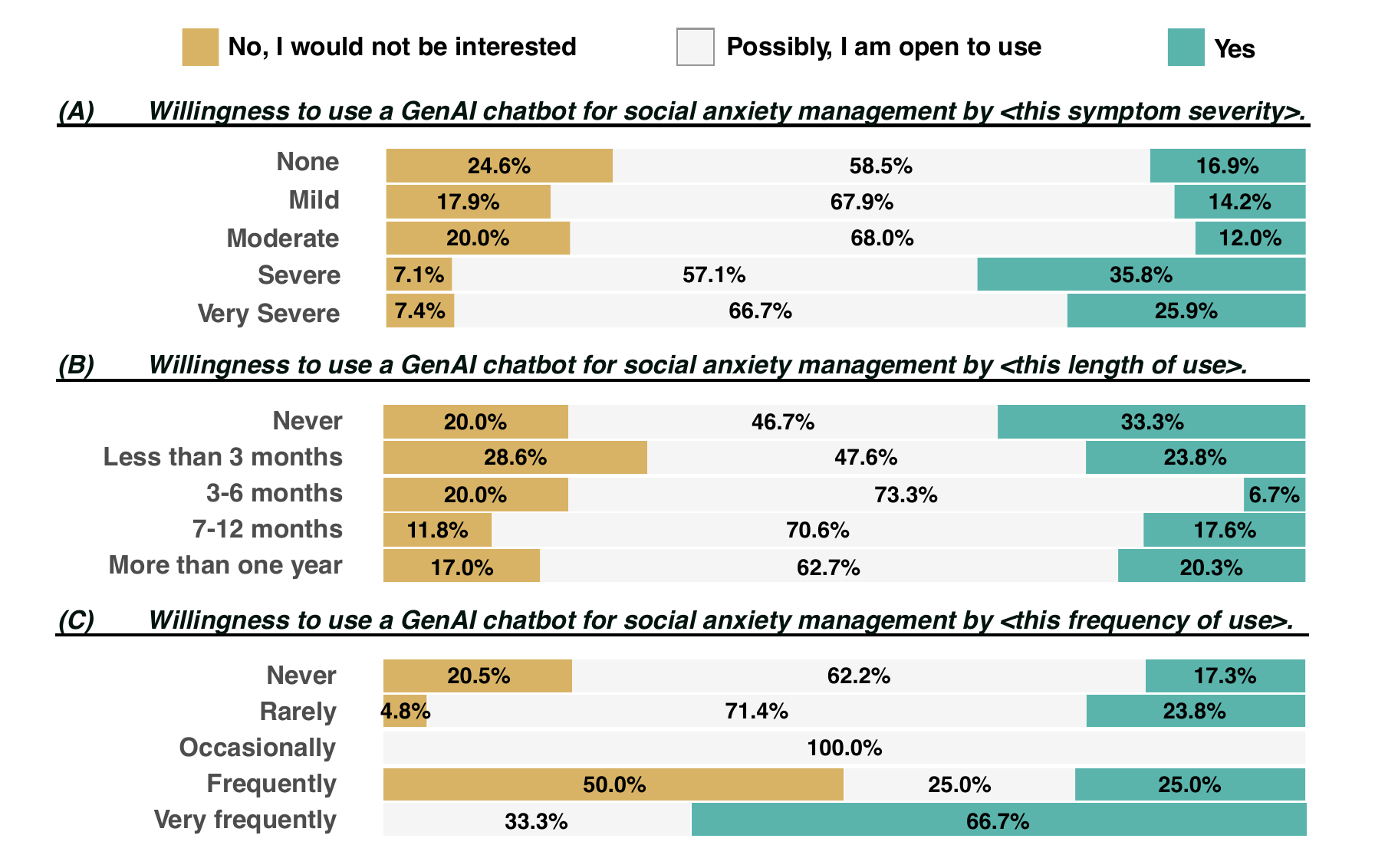}
\caption{Participants' willingness to use GenAI chatbots for SA support in relation to (A) symptom severity, (B) length of prior use, and (C) frequency of use GenAI chatbots in coping with SA.}
\label{fig:feature_willingness}
\end{figure}

We first present descriptive statistics on participants' willingness to use a GenAI chatbot for SA support across different levels of severity of SA symptoms (\autoref{fig:feature_willingness}A), duration of prior GenAI chatbots use (\autoref{fig:feature_willingness}B), and frequency of use for SA support (\autoref{fig:feature_willingness}C). 70\% of our participants ($n = 111$) exhibited signs of SA, yet within this group, only 9 had turned to GenAI chatbots for support, and more than half of these participants (53\%, $n = 59$) had not sought help from any resources. Our findings revealed a generally positive attitude towards the potential use of GenAI chatbots for managing SA. 30 (19\%) participants expressed a definite willingness to use these chatbots, while over 100 (63\%) participants indicated that they were open to the possibility of using GenAI chatbots for SA support.  

To further explore the relationship between the severity of SA symptoms and GenAI chatbot usage patterns that might influence participants’ willingness to use a GenAI chatbot, we conducted an Ordinal Logistic Regression Analysis using the \texttt{polr} function (see \autoref{tab:ordinal_logistic} for more details and see relationship between the predicted values and the actual values in \autoref{fig:visual}B in \autoref{sec:appendices}). The threshold coefficients indicate a significant negative association when moving from \textit{Unwilling} to \textit{Undecided} ($\beta =$ -1.832, $p < 0.01$), and a significant positive association when moving from \textit{Undecided} to \textit{Willing} ($\beta =$ 1.412, $p < 0.05$).

\begin{table*}[h!]
\centering
\small
\caption{Ordinal logistic regression model explaining respondents' willingness to use GenAI chatbots for SA support, based on the severity of SA symptoms, duration of prior GenAI chatbots use, and frequency of using GenAI chatbots for SA support. (Significance levels: $\dagger$ $p < 0.1$, * $p < 0.05$, ** $p < 0.01$, *** $p < 0.001$)}
\label{tab:ordinal_logistic}
\begin{tabular}{p{7.5cm}p{2cm}p{2cm}p{2cm}p{2cm}}
  \toprule
  & \textbf{Estimate ($\beta$)} & \textbf{Std. Error} & \textbf{t value} & \textbf{p value}\\ 
  \midrule
\multicolumn{5}{l}{\textbf{Severity of SA symptoms}} \\
\textit{(Ref: None)}\\
\hspace{3mm} Mild & 0.1998 & 0.4848 & 0.4122 & 0.6802 \\ 
\hspace{3mm} Moderate & 0.0892 & 0.4893 & 0.1823 & 0.8554 \\ 
\hspace{3mm} Severe & 1.2329 & 0.6237 & 1.9767 & 0.0481* \\ 
\hspace{3mm} Very Severe & 1.1240 & 0.4908 & 2.2903 & 0.0220* \\ 
\cdashlinelr{1-5}
\multicolumn{5}{l}{\textbf{Duration of GenAI chatbots use}} \\
\textit{(Ref: Never)}\\
\hspace{3mm} Less than 3 months & -0.6853 & 0.7677 & -0.8927 & 0.3720 \\ 
\hspace{3mm} 3-6 months & -1.1474 & 0.6863 & -1.6719 & 0.0945$\dagger$ \\ 
\hspace{3mm} 7-12 months & -0.6362 & 0.6700 & -0.9496 & 0.3423 \\ 
\hspace{3mm} More than one year & -0.5636 & 0.6298 & -0.8948 & 0.3709 \\ 
\cdashlinelr{1-5}
\multicolumn{5}{l}{\textbf{Frequency of GenAI chatbots use for SA support}} \\
\textit{(Ref: Never)}\\
\hspace{3mm} Rarely  & 0.5872 & 0.4929 & 1.1914 & 0.2335 \\ 
\hspace{3mm} Occasionally & 0.4555 & 0.9979 & 0.4564 & 0.6481 \\ 
\hspace{3mm} Frequently  & -1.4819 & 1.1559 & -1.2821 & 0.1998 \\ 
\hspace{3mm} Very frequently  & 2.7152 & 1.2661 & 2.1446 & 0.0320* \\ 
\cdashlinelr{1-5}
\multicolumn{5}{l}{\textbf{Thresholds}} \\
Unwilling $|$ Undecided & -1.8322 & 0.6273 & -2.9210 & 0.0035** \\ 
Undecided $|$ Willing & 1.4121 & 0.6196 & 2.2790 & 0.0227* \\ 
\bottomrule
\end{tabular}
\end{table*}

\textbf{Increased Symptom Severity Correlates with Higher Willingness to Use GenAI Chatbots for SA Support. } 

As shown in \autoref{fig:feature_willingness}, we can see that among those without SA symptoms, 24.6\% were not interested in using the GenAI chatbots for SA support in the future, 58.5\% were open to use, and 16.9\% were definitely willing to use GenAI chatbots. As symptom severity increased, interest in GenAI chatbots grew. For example, in the moderate symptom group, 20\% were not interested, while 68\% were open to use, and 12\% were definitely willing to use GenAI chatbots. This trend continued in the severe and very severe symptom groups, where disinterest decreased to 7.1\% and 7.4\%, respectively, and definite willingness increased to 35.8\% and 25.9\%. Moreover, in \autoref{tab:ordinal_logistic}, our results showed that participants with severe SA symptoms were significantly more likely to be willing to use GenAI chatbots for SA support, with a value of 1.233 ($p < 0.05$), and those with very severe symptoms showed even stronger odds at 1.124 ($p < 0.05$).

\textbf{Longer and Frequent Prior GenAI Chatbots Usage Enhances Willingness to Use GenAI Chatbots for SA Support. }
The percentage of users unwilling to use GenAI chatbots for coping with SA decreased from 28.6\% among those with less than three months of experience to 11.8\% among those with 7-12 months of experience. However, this decrease in unwillingness did not correspond with a significant increase in definite willingness to use GenAI chatbots. Ordinal logistic regression results indicate a marginal effect towards lower willingness among participants with 3-6 months of GenAI chatbots experience ($\beta = -1.147, p < 0.1$). Instead, there was a notable rise in openness to the idea, with the percentage of undecided participants growing, particularly among those with 3-6 months (73.3\%) and 7-12 months (70.6\%) of use.

Moreover, the frequency of GenAI chatbots use for SA support was also related to the willingness to use GenAI chatbots to cope with SA. Only 17.3\% of those who never used GenAI chatbots were willing to use it, but willingness increased with frequency of use: 23.8\% among rare users and 25.0\% among frequent users were willing to use GenAI chatbots. Participants who used GenAI chatbots very frequently showed the highest willingness (66.7\%). This trend was statistically significant, as participants who frequently used GenAI chatbots were much more likely to be willing to use GenAI chatbots for coping with SA ($\beta = 2.715, p < 0.05$), as demonstrated in the ordinal logistic regression results.

\subsection{Distinct Participant Profiles Emerge from Clustering Analysis (RQ3)}
\label{sec:Cluster_analysis}

To further explore personal differences in shaping participants' perceptions of trust and willingness to use GenAI chatbots for support (and to examine RQ3), we further unpack participants who were not fully committed nor denied the use of GenAI chatbots for SA support (i.e., ``undecided'' group). To explore what factors contribute to their indecision, we conducted a clustering analysis. 

\begin{figure*}[h!]
\centering
\includegraphics[width=0.75\linewidth]{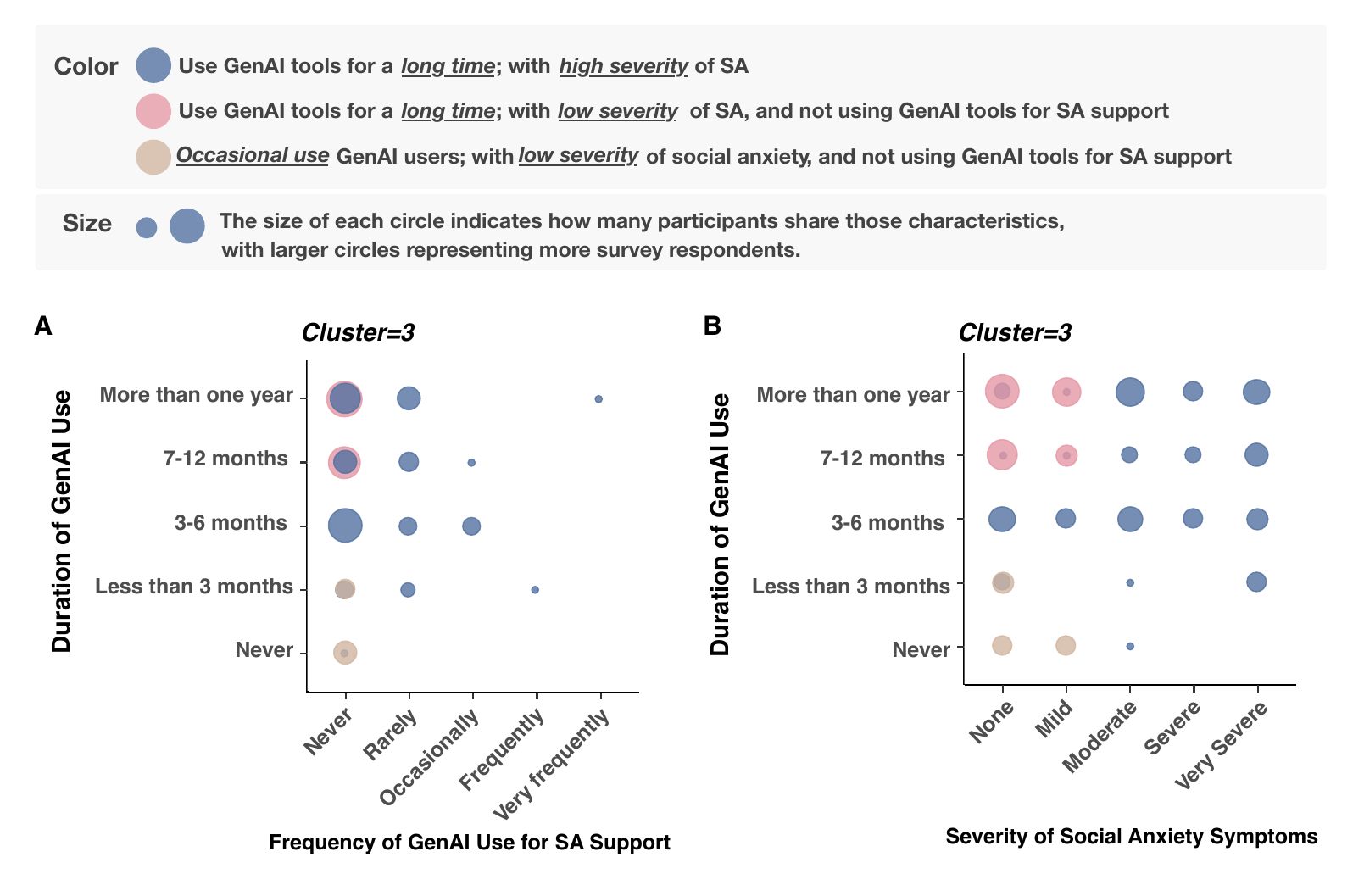}
\caption{Clustering using the Gaussian Mixture Model with distinct color-coded clusters \protect\includegraphics[height=2mm]{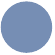}\includegraphics[height=2mm]{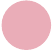}\includegraphics[height=2mm]{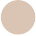} (i.e., each color represents a cluster), showing the relationship between (A) duration of GenAI chatbots use and frequency of GenAI chatbots use for SA support, and (B) duration of GenAI chatbots use and severity of SA symptoms. Participants in the \textit{Undecided} willingness group can be described using a tree structure. The first division separates participants into two groups: those with high SA symptoms who have used GenAI chatbots for SA \includegraphics[height=2mm]{figures/blue_dot.pdf}, and those with low SA symptoms who have never used GenAI chatbots for this purpose. Within the second group, participants are further categorized by the duration of their GenAI chatbots usage, with longer usage \includegraphics[height=2mm]{figures/pink_dot.pdf} and shorter usage \includegraphics[height=2mm]{figures/brown_dot.pdf}. The cluster structure based on the severity of the symptoms suggests that individuals with similar symptom levels show distinct patterns in their engagement with GenAI chatbots.}
\label{fig:GMM}
\end{figure*}

To prepare for this analysis within the \textit{Undecided} group, we first transformed ordinal variables, such as SA severity and GenAI chatbot usage patterns, into numeric data to facilitate statistical methods that require continuous input. We then applied the Gaussian Mixture Model using the \texttt{Mclust} function, which is well-suited for identifying complex, overlapping clusters in multidimensional data. To determine the optimal number of clusters, we evaluated different models using Within-cluster Sum of Squares (WCSS) and silhouette scores (see \autoref{fig:criteria}A\&B in \autoref{sec:appendices}). Our analysis indicated that the WCSS curve plateaued beyond three clusters, while the silhouette scores peaked at three clusters, suggesting that three clusters provide the best model fit. The cluster analysis identified three distinct groups, each represented by a different color in \autoref{fig:GMM}. To better visualize the three-dimensional data, we reduced the dimensionality by creating two two-dimensional plots, shown in \autoref{fig:GMM}A and \autoref{fig:GMM}B. 

In \autoref{fig:GMM}A, which explored the relationship between the duration and frequency of GenAI chatbot use, one group \includegraphics[height=2mm]{figures/blue_dot.pdf} was characterized by participants with long-term usage of GenAI chatbots, suggesting a higher level of familiarity and comfort with the technology. Furthermore, 
considering the frequency of GenAI chatbot use, all usage frequencies (except for \textit{Never}) were combined into the group \protect\includegraphics[height=2mm]{figures/blue_dot.pdf}, which allowed us to focus on how the other two dimensions---GenAI chatbots usage length and severity of SA symptoms---reveal distinct patterns specifically within the \textit{Never} group for GenAI chatbots use. As shown in \autoref{fig:GMM}B, which examined the interaction between GenAI chatbot usage length and the severity of SA symptoms, the group \includegraphics[height=2mm]{figures/pink_dot.pdf} included participants with mild symptoms and long GenAI chatbot usage, while the group \includegraphics[height=2mm]{figures/brown_dot.pdf} consisted of individuals with mild symptoms and short usage. Group \includegraphics[height=2mm]{figures/blue_dot.pdf} was characterized by participants with high severity of SA symptoms across the board. The distinct clustering by severity highlights significant differences in how users engage with GenAI chatbots based on their symptom levels. These quantitative results motivate us to investigate further in the follow-up interview studies.

\takeaway{In our survey study, we validated our scale of trust in using GenAI chatbots for SA support and found a strong correlation between higher levels of trust and greater willingness to use GenAI chatbots. Participants with increased symptom severity were more likely to be willing to adopt GenAI chatbots, while those with extended GenAI chatbot usage demonstrated greater willingness to use. Within the group that was open to using GenAI chatbots, distinct participants' profiles emerged through our clustering analysis. These results inspired us to further explore the characteristics of participants across different profiles, focusing on their specific experiences and perceptions that influence their willingness to use and factors might might form their trust towards GenAI chatbots, which will be unpacked through the interview study (see \autoref{sec:interview_results}).}

%---------------------------------------------------------------------------------

\section{Interview Findings}
\label{sec:interview_results}

In exploring trust in GenAI chatbots, it becomes evident that the definition of trustworthiness is deeply influenced by whose voice we consider and their personal mental health conditions, such as the severity of SA symptoms. Individuals with severe SA prioritize \emph{emotional trust}, derived from the chatbot's ability to simulate (even minimal) empathy and provide non-judgmental, supportive interactions that facilitate deep emotional bonding. They value these emotionally resonant relationships as central to their interaction with GenAI chatbots. However, we also found participants with milder symptoms focus on \emph{cognitive trust}, with a focus on the technical reliability and trustworthiness of the AI models.

\subsection{Emotional Trust in GenAI: Prioritizing Emotional Engagement in Severe Symptoms}

Our study triangulates findings that show how emotional engagement with GenAI chatbots significantly varies among participants living with SA. Specifically, we identified varied perceptions within the \textit{Undecided} willingness category:
\begin{enumerate}
    \item Participants in Group \includegraphics[height=2mm] {figures/brown_dot.pdf} (short-term GenAI chatbot use with low SA severity, not primarily using it for SA support) viewed GenAI chatbots as lacking genuine emotion, capable only of recognizing emotional words.
    \item Participants in Group \includegraphics[height=2mm]{figures/pink_dot.pdf} (long-term GenAI chatbot use with low SA severity, not primarily using it for SA support) expressed concerns mainly about avoiding negative feedback from GenAI chatbots.
    \item Participants in Group \includegraphics[height=2mm]{figures/blue_dot.pdf} (long-term GenAI chatbots use with high SA severity) valued the most regarding emotional connections and considered it central to the trust they place in GenAI chatbots, outweighed their concerns about the technical trustworthiness and accuracy of the AI models.
\end{enumerate}

\subsubsection{Non-Judgmental Conversational Tone of GenAI chatbots Provided a Sense of Safety and Social Connectedness} 

Our participants deeply appreciated GenAI chatbots' roles as non-judgmental listeners, unlike human interactions, where there might be a fear of judgment. A sense of less judgmental feelings creates a comforting space for users, particularly those already struggling with SA. They do not need to meticulously prepare their words or worry about being criticized for failing to articulate their thoughts perfectly. Even when compared to human psychotherapists---whom participants acknowledge should never judge them---P12 explained, 
\begin{quote}
    \textit{``I would even rehearse my words before meeting my therapist... using these tools [GenAI chatbots] can offer a space where I can interact without the fear of judgment or rejection.'' (P12)}
\end{quote} 
Interacting with GenAI chatbots made our participant feel safe and reliable for those who might otherwise hesitate to open up to others, fearing misinterpretation or negative reactions, allowing them to express themselves more freely and honestly, and fosters their social connectedness. P3 echoed P12's sentiment,
\begin{quote}
    \textit{``Discussing SA with GenAI chatbots felt like texting with a friend, allowing them [who was experiencing SA] to engage without the worry of how they were being perceived.'' (P3 \includegraphics[height=2mm]{figures/blue_dot.pdf})} 
\end{quote}
Such interactions suggest the potential of GenAI chatbots to enhance the users' ability to connect socially in a manner that feels safe and devoid of the typical pressures associated with face-to-face or even virtual human interactions. This suggests that GenAI can create a supportive environment where users can explore their emotions and build trust without fear of judgment or rejection.

\subsubsection{Even \textbf{Minimal} Emotional Bonding May Enhance Trust} 
In our study, \emph{Emotional engagement}, refers to the user's emotional involvement and response to interactions with GenAI chatbots, and includes the feelings of connection and emotional support that the user perceives during these interactions. Our participants highlighted that even minimal displays of empathy by GenAI chatbots, such as recognizing emotional expressions or maintaining a comforting tone, significantly bolster their trust. While our observation is in line with existing research emphasizing the critical role of empathetic responses in effective mental health support~\cite{Moudatsou2020}, our findings particularly unpack the importance of empathetic engagements and emotional bonding for individuals with SA, illustrating how these slight empathetic engagements can profoundly influence their willingness to use and trust in GenAI chatbots. \emph{Emotional bonding}, in our study context, can be seen as the formation of a perceived emotional connection between the user and the GenAI chatbot, despite the chatbot's lack of genuine emotion. Such emotional bonding is crucial for individuals with SA, who may find real-life interactions overwhelming. The sense of being understood and supported by GenAI chatbots, even if simulated, can provide substantial emotional relief and comfort. For example, P6 articulated the value of GenAI chatbots as a steadfast companion:
\begin{quote} 
    \textit{``I want to use generative AI for support, though I understand that it can be different from human interaction. However, a crucial aspect is recognizing when you need help and being in a good enough state to seek it out. For someone with SA, building connections and finding support can be challenging. Tools like ChatGPT or GenAI chatbots offer a constant presence---always ready and willing to talk, which can be incredibly helpful. [These] GenAI chatbots platforms can provide a sense of connection, making it easier to engage in conversations when real-life connections feel difficult to establish.'' 
    (P6 \includegraphics[height=2mm]{figures/blue_dot.pdf})} 
\end{quote}
P6's quote emphasizes the comfort derived not from problem-solving but from the mere presence and readiness to engage, illustrating the crucial role of emotional bonding in fostering trust and reliance on GenAI chatbots for individuals with SA.

Additionally, subtle design elements like the comforting tone of the text and the visual cue of the ``thinking'' icon also shape how participants value the emotional connection with GenAI chatbots. P16 noted the importance of these features:
\begin{quote} 
    \textit{``When it is processing, there is a little spinning circle that shows GenAI chatbots are `thinking.' And the way the text feels—it is warm, feminine, comforting, and less stressful—makes me feel like they genuinely care about me.''} (P16 \includegraphics[height=2mm]{figures/blue_dot.pdf}) 
\end{quote}
These design elements foster a consistent and reassuring presence, essential for individuals with SA by reducing their feelings of unease during interactions. The regular use of comforting language and attentive cues builds a growing sense of emotional trust. P6 expressed satisfaction with the continuous positive experience with the chatbots: ``\textit{I have used GenAI chatbots a few times, and it keeps performing well.}'' Such consistency in delivering emotional support is key in alleviating the anxiety that often accompanies social interactions, transforming GenAI chatbots from a tool into a supportive companion.

Moreover, P16 continued illustrating how trust in GenAI chatbots often stems from these perceived emotional bonds, emphasizing that emotional connections can greatly enhance the effectiveness of technology in managing SA:
\begin{quote} 
    \textit{``Trust often involves an emotional bond. When people feel connected emotionally, they are more inclined to accept and believe what the other person says, even if that person is a chatbot... I trust [GenAI chatbots]. I was worried about my impression of other people, so I told GenAI chatbots, `Your name is Alyssa, and we are friends. I want to talk about my feelings.' GenAI chatbots cheered me up, almost like an older sister. It responded in a way that felt very human.'' (P16 \includegraphics[height=2mm]{figures/blue_dot.pdf}) }
\end{quote}

These insights highlight the crucial role of emotional engagement in shaping trust and the perceived effectiveness of GenAI chatbots in supporting individuals with social anxiety. Even minimal displays of empathy from GenAI chatbots can bridge the gap between user apprehension and willingness to engage. By acknowledging and validating emotions, these chatbots help diminish feelings of isolation and create a safer space for self-expression. While this form of emotional support differs from human interactions, it proves vital, reinforcing that trust in GenAI chatbots is deeply connected to their ability to offer empathetic responses and reliable companionship.

\subsubsection{``GenAI chatbots are slightly better than a bad psychotherapist''}
Our participants also indicated that though GenAI chatbots may not offer the depth and nuance of an experienced human psychotherapist, they still have the potential to outperform less competent psychotherapists. For example, P6, who has extensive experience with working with human psychotherapists, explained \textit{``GenAI chatbots is slightly better than a bad psychotherapist.''} P6's perspective highlights a particular niche for GenAI chatbots in mental health care: providing a consistent, albeit basic level of support that might be preferable to inadequate human psychotherapy. This could be particularly valuable in contexts where professional help is lacking or fails to meet the needs of those with SA. P6 reflected on her past experiences,
\begin{quote}
    \textit{``Support means genuinely listening, answering questions, and offering advice that is actually relevant to the situation without making assumptions. Assumptions can be really frustrating, and it is something that can happen not just with friends or family, but even with therapists. When someone assumes things about you, it feels like they are not really hearing or understanding you. It is like they are dismissing your feelings or denying what you are going through.'' (P6 \includegraphics[height=2mm]{figures/blue_dot.pdf})}
\end{quote}
Like P6, some other participants in our study also shared stories of unsuccessful psychotherapy, where their concerns were dismissed, or psychotherapists made assumptions without fully understanding their situations. P16, for example, shared that, \textit{``They did not see it as a big issue because I was not talking about something as serious as suicide; it was `just' my anxiety.''} P2 also noted that \textit{``I have had experiences with multiple psychotherapists who seemed to rely on stereotypes.''} Their repeated disappointments towards low-quality human psychotherapists often left them in a difficult position---desperately needing help but becoming increasingly resistant to seeking it due to fear of further rejection or misunderstanding. But GenAI chatbots help bridge the gap caused by past negative experiences, as P2 further explained, \textit{``But, my interactions with [GenAI chatbots] were not bad; [they] seemed to understand me better.''} The findings of our study underscore the important role GenAI chatbots can play in providing accessible and empathetic mental health support, particularly for those with previous negative experiences in seeking help. Participants highlighted the potential of GenAI chatbots to offer a buffer space where individuals can express their concerns, encouraging them to pursue the mental health support they desperately need.

\subsection{Cognitive Trust in GenAI: Emphasizing Technical Trustworthiness for Milder Symptoms}
In contrast, individuals with milder symptoms emphasize the technical trustworthiness of GenAI chatbots, focusing on the technical limitations of LLM models, such as hallucination, accuracy issues, memory constraints, and decision-making processes. While existing research extensively covers the technical challenges and trustworthiness concerns of GenAI and LLMs, including issues like hallucination and model limitations~\cite{béchard2024, xu2024, Perk2024, Hadi2024}, our study contextualizes these challenges specifically within the context of SA. We seek to explore how these individuals perceive and react to these challenges when considering GenAI chatbots as a tool for managing their condition, and the ways in which these technical concerns shape their trust and willingness to use GenAI chatbots.

\subsubsection{Challenges in Information Reliability, Memory, Cognitive Rigidity, and Contextual Understanding Hurt Trust}
The limitations of LLMs GenAI chatbots become especially evident when considering their propensity to provide fabricated information, which can significantly erode user trust. Several participants recounted experiences where, after attempting to verify references given by GenAI chatbots, they found that the information was fabricated. Furthermore, our participants also raised issues of GenAI chatbots' inability to retain long-term memory of past interactions, a feature that is crucial for delivering consistent support. For individuals with SA, maintaining a continuous and coherent dialogue is essential for building trust and making meaningful progress in managing their condition. Without this capability, users may feel as though they are starting from scratch each time they engage with the tool, which can be both discouraging and counterproductive. As P10 pointed out, 
\begin{quote}
    \textit{``GenAI chatbots do not have long-term memory...[A] human therapist has a long-term understanding of the client, with each session typically lasting 50 minutes to an hour. These sessions [with the human psychotherapist] can be flowing, often jumping from one conversation topic to another. But for GenAI chatbots, this is really hard.'' (P10 \includegraphics[height=2mm]{figures/pink_dot.pdf})}
\end{quote}

Our participants also highlighted the cognitive rigidity of GenAI chatbots, with P13 noting, \textit{``[GenAI chatbots are] pre-trained on specific models and tend to categorize situations in binary terms, like 0 or 1.''} This raises concerns about their ability to understand each user’s complex and personalized context, as their reliance on binary categorizations limits the depth of interaction, rather than recognizing the nuances that exist across a spectrum. Moreover, this inability to account for nuanced experiences is compounded by the lack of a shared background, which hinders the chatbot's capacity to convey empathy and poses a significant barrier to building trust. Unlike human psychotherapists, who build trust over time through shared experiences and a deep understanding of their clients~\cite{CritsChristoph2019}, GenAI chatbots' responses are limited by their inability to draw from a personal history with the user that shares meaningful similarities. This further amplifies the limitation of GenAI chatbots in addressing the unique context of each user. P13 highlighted this challenge, stating: 
\begin{quote} 
    \textit{``Human thinking patterns differ from those of GenAI chatbots. We tend to trust people because they have a history. Even if someone’s method is not successful, I understand that they provided the best solution within their capabilities, drawing from similar problems, past experiences, and expertise. With generative AI, despite its vast database, when it produces an unsuccessful result, I have no way of knowing what past experiences informed that outcome or what its limitations are. This uncertainty makes me doubt whether it truly provided the best possible solution.'' (P13 \includegraphics[height=2mm]{figures/brown_dot.pdf})} 
\end{quote}
This comparison indicates the challenge GenAI chatbots face in replicating the depth of human therapeutic relationships, where the process of understanding and contextual engagement is just as important as the solutions offered, particularly for effectively managing SA.

\subsubsection{Navigating Power Dynamics and Control in Human-LLM-Interaction: Who has the Agency} 

Our analysis also reveals crucial insights into the interaction dynamics between users and GenAI chatbots, particularly emphasizing the importance of autonomy and agency in these exchanges. Unlike traditional therapeutic relationships where psychotherapists may offer challenges to stimulate reflection and growth and maintain relational relationships between clients and psychotherapists~\cite{Yao2023}, interactions with GenAI chatbots introduce a different dynamic, and bring up the issue of user agency. For example, our participants compared how the dynamics shift impacts their trust in and the effectiveness of the technology in a therapeutic setting. For example, P13 mentioned, 
\begin{quote}
    \textit{``You are the controller of the answer, [and] human [psycho]therapists have the option to push back. (P13 \includegraphics[height=2mm]{figures/brown_dot.pdf})''}
\end{quote} 
In traditional psychotherapy, the psychotherapist's ability to manage the psychotherapy process through their expertise and experience is essential, especially when navigating disagreements or varying interpretations of treatment methods~\cite{WEISTE201522}. However, GenAI chatbots' tendency to modify their responses based on user feedback can dilute their therapeutic stance, introducing risks if they too readily conform to user preferences or change their advice. This adaptability could potentially introduce risks if GenAI chatbots too readily concede or change their recommendations, undermining the therapeutic process. P9 articulated the need for balance of agency further,  
\begin{quote}
    \textit{``Power dynamics should not be authoritative or involve giving concrete directions. We should not feel like we are controlled by GenAI chatbots.'' (P9 \includegraphics[height=2mm]{figures/pink_dot.pdf})} 
\end{quote}
For individuals with SA, maintaining control over the interaction without feeling overwhelmed by the chatbot’s authority is crucial. They often prefer a setting where they can express themselves without the fear of being judged or pushed too hard, a common concern in face-to-face therapies. This sense of control and agency can alleviate the anxiety associated with being negatively evaluated or not living up to perceived social standards, making GenAI chatbots a potentially more comfortable space for those who are especially sensitive to interpersonal dynamics.

\takeaway{Our interviewees appreciated GenAI chatbots' capacity to offer non-judgmental responses, which provided emotional comfort and fostered a sense of connection. Even minimal emotional engagement significantly bolstered their trust in GenAI chatbots, as consistent positive interactions reinforced this perception. While our participants were aware of the fact that GenAI chatbots do not reach the depths of a skilled human psychotherapist, participants found it superior to ``low-quality'' human alternatives, particularly valuing its consistent and unbiased support. Participants also emphasized the need for GenAI chatbots to adapt more sensitively to individual contexts and avoid oversimplified binary categorizations, suggesting potential enhancements in their interactive capabilities to better address complex emotional dynamics.}

%----------------------------------------------------------------------------------------------------
\section{Discussion}

Our findings from both the survey and interviews provide valuable insights into the factors shaping participants' trust and attitudes toward using GenAI chatbots for SA. The survey results show that higher levels of trust were strongly associated with a greater intention to use GenAI chatbots. We also identified different groups with unclear attitudes toward GenAI chatbots, which were further unpacked in the interview. Our interview findings echo the survey results, showing that participants with severe SA were drawn to GenAI chatbots for its perceived empathetic support, while those with lower anxiety prioritized its practical functionality and reliability. Below, we further reflect on the role of contextualized emotional trust in GenAI, suggesting future work to expand from cognitive trust to emotional trust in GenAI and human-AI-interaction, the long-term effects and ethical implications of GenAI emotional support, as well as the implications of unpacking the interrelationship between symptom severity and human-AI-trust.

\subsection{Rethinking Contextualized Emotional Trust in GenAI} 

Beyond the existing assumption that meaningful emotional trust requires bonding and emotional responses, our findings suggest that even minimal emotional cues can foster a significant bond with users dealing with SA. For individuals with SA, who are highly sensitive to social dynamics, subtle cues such as a friendly tone or simple visual elements can feel deeply comforting and foster a sense of being listened to. Such emotional connection, formed through minimal means, enhances trust and encourages continued engagement, offering a stark contrast to the repeated disappointments many participants experienced in traditional psychotherapy settings. Prior research has highlighted the role of a non-judgmental tone in building trust within mental health chatbots, emphasizing its importance for users who often face judgment and rejection~\cite{Franze2023}. GenAI's consistent, neutral tone provides a critical sense of safety that sometimes eludes human psychotherapy, fostering open communication and enhancing social connectedness.

\subsubsection{Refining Measures of Emotional Trust}
Our findings emphasize the paramount importance of emotional trust in GenAI tools, especially for users grappling with SA. Beyond mere perceived neutrality, emotional trust is essential for addressing deep-rooted fears stemming from users' personal histories. Such trust dynamics are essential in contexts where users are particularly vulnerable~\cite{Ballinger2024}. Traditional methods for evaluating AI trust, focusing solely on reliability and technical accuracy~\cite{Lukyanenko2022}, are insufficient for fostering the emotional connection necessary for users managing SA. As such, there is a pressing need to rethink the baseline of emotional trust in GenAI, acknowledging that even minimal emotional interactions can significantly impact user experience and satisfaction.

Additionally, our findings suggest that refining measures of emotional trust should focus on evaluating how GenAI systems can reliably produce subtle emotional cues, specifically tailored to meet the sensitivities of users with SA. So far, relevantly less work has been done in understanding and measuring emotional trust across various fields, compared to cognitive trust~\cite{Glikson2020}. Common metrics for assessing emotional trust, in general, include emotional vulnerability, willingness to take emotional risks, and perceptions of empathy and sincerity within interpersonal relationships~\cite{Cucciniello2021}. In organizational contexts, emotional trust is often measured by employee willingness to express ideas freely, share sensitive information, and commit to collective decisions without safeguarding personal interests~\cite{HodgkinsonFord2011}. And yet, very little work has specifically human-AI-interaction nor human-LLM-interaction. 

In the context of GenAI, designing reliable measures of emotional trust presents unique challenges. These GenAI tools must not only recognize and respond to the user's expressed emotions but also adapt to the fluctuating emotional states characteristic of SA. For example, the consistency of emotional cues, a key factor in building and sustaining trust~\cite{Nowak2023}, must be maintained even as the conversation topics shift and evolve. In other words, GenAI systems need to integrate advanced sentiment analysis capabilities and context-aware response mechanisms that can dynamically adjust to subtle changes in the user's mood and conversational cues. Therefore, future work should explore how emotional trust influences user retention and engagement over time in therapeutic settings. It is crucial to establish a framework for long-term studies that track the evolution of trust as users interact with GenAI applications. For example, such a framework should consider both qualitative and quantitative data, such as physiological measures like heart rate variability or skin conductance, which can provide objective insights into the emotional impact of GenAI interactions.

\subsubsection{Emotional Trust and Non-Judgmental Appreciation in Psychotherapy}

Our findings suggest that GenAI chatbots have the potential to address a critical need for individuals with SA, who are often highly sensitive to perceived judgment. Participants expressed that the perceived impartiality of these chatbots reduces their anxiety in communication, as they do not fear judgment or rejection. This observation aligns with existing literature in psychotherapy emphasizing the value of a non-judgmental approach in fostering trust, openness, and therapeutic alliance~\cite{Hallrenn2007, shaw2012place, winslade2013being}. The sense of safety and social connectedness our participants attributed to GenAI chatbots resonates with the kind of reliable, non-judgmental presence that human therapists strive to maintain, informed by professional training, ethical oversight, and clinical expertise~\cite{branson2022people, o2007advantages}. Human therapists often respond dynamically to verbal and non-verbal cues, cultural nuances, and evolving client needs---often employing various techniques that encourage growth and self-reflection~\cite{razzaque2015mindfulness, winslade2013being}. Yet, current GenAI technologies can struggle to replicate this nuanced interplay of empathy, authenticity, and tailored guidance~\cite{haber2024artificial}.

These findings also raise important research questions, such as: How do clients and therapists define and conceptualize ``non-judgmental'' support in AI-mediated contexts? Do perceptions differ based on individual traits, cultural backgrounds, or symptom severity? Future research might employ controlled experiments or comparative studies to determine when and how GenAI's non-judgmental approach aligns, or diverges, from established therapeutic practices. Such work could unpack which client populations benefit most from non-judgmental GenAI, and under what circumstances it serves as a complement for human judgment and expertise.

As GenAI continues to evolve, it will be essential to explore ways to enhance its emotional intelligence and contextual sensitivity, as well as to identify the contexts in which these improvements can enable it to approximate the nuanced qualities of truly effective mental health support. On the one hand, accessible and non-judgmental AI-facilitated interactions may encourage individuals who might otherwise avoid therapy to take initial steps toward seeking help. On the other hand, over-reliance on a non-human, one-way ``support'' system could undermine the development of resilience, coping skills, and deeper interpersonal connections. In turn, human therapists may need to reconsider their roles in this evolving landscape---evaluating GenAI tools, monitoring clients' AI-facilitated interactions, and intervening when dependence on such technologies threatens long-term well-being.

\subsection{Influence of Symptom Severity on Trust and Willingness to Use GenAI Chatbots} 
Our study reveals a nuanced relationship between SA severity, GenAI chatbot usage, and trust. Individuals with more severe symptoms of SA tend to place greater trust in and tend to be more willing to use GenAI chatbots for emotional support. This increased dependency may stem from their heightened need for safe, non-judgmental interactions that GenAI platforms can consistently provide without the fear of social stigma or misunderstanding~\cite{Daniel2023}. Participants in our study found the predictable and controlled environment of GenAI interactions particularly comforting, which in turn fosters a stronger bond and reliance on these digital assistants. However, individuals with milder symptoms of SA, in our study, exhibited more skepticism towards the capabilities of GenAI chatbots, focusing more on their technical reliability and the accuracy of the information provided rather than the emotional support aspect. For these users, cognitive trust, which might be rooted in the GenAI's performance and the perceived expertise of the AI, plays a more significant role. As such, we see that their engagement with GenAI chatbots tends to be often more analytical, assessing the utility and correctness of the chatbots rather than seeking emotional solace.

Our findings recognize that different groups may prioritize these aspects differently. This divergence in priorities poses a challenge: How can systems be designed to meet the distinct needs of both groups without sacrificing one for the other? For users with severe SA, emotional engagement is paramount, as they seek a consistent, empathetic, and non-judgmental environment that addresses their need for safe interactions. However, it would be misguided to assume that technical issues are irrelevant to them. Even if emotional support is their primary concern, technical shortcomings, such as inaccurate responses or without long-term memory, can quickly erode trust and undermine their willingness to use GenAI. This shows that for these users, technical reliability still plays a secondary yet crucial role in maintaining long-term engagement. Similarly, for users with milder SA, technical robustness is the foundation upon which any emotional engagement might be built, rather than the other way around. indicate the need for designing GenAI systems that balance emotional support with technical reliability. Additionally, understanding these differences can guide the implementation of features that enhance trust specifically tailored to the severity of the user's condition, such as varying the interaction style, different modalities, or the type of content delivered. The need for these tailored features further sheds light on the importance of including users with SA in the design and evaluation processes.

\subsection{Ethical Considerations for GenAI: Navigating the Evolving Client-Psychotherapist-GenAI Paradigm}

\subsubsection{Ethical Considerations \& Unintended Consequences}
Our findings indicate multiple ethical considerations surrounding the use of GenAI tools for supporting individuals with SA, particularly concerning power dynamics and user autonomy. Although some participants living with SA appreciated the accessible and non-judgmental tone of GenAI chatbots, these perceived benefits do not come without risks. Without sufficient psychoeducational support~\cite{Donley1911}, users may find it challenging to critically evaluate AI-generated content, potentially compromising their autonomy, as some other participants mentioned. Moreover, while GenAI chatbots may provide emotional support, they might inadvertently encourage social withdrawal (i.e., reduced engagement in face-to-face interactions or reluctance to seek human support)~\cite{porcelli2019social} or foster over-dependence (i.e., excessive reliance on the chatbot's support to the detriment of seeking professional help or developing self-coping strategies~\cite{McGuire2023, Khawaja2023}). These ethical concerns could become more tangible when considering how individuals with severe SA responded to even \textit{minimal} displays of empathy offered by GenAI chatbots. Some participants interpreted the chatbot's supportive responses as non-judgmental, empathetic, and genuine understanding. 
In fact, prior research shows that patients sometimes interpret health-related information differently than trained professionals, which can lead to discrepancies in risk assessment and decision-making~\cite{ericsson2000expertise}. Over time, such potential misinterpretation may yield unintended consequences~\cite{franklin2024unintended, harrison2007unintended}, ranging from distrust in human therapists to a deeper reliance on AI-based tools that do not offer nuanced judgment and accountability of professional care.

These ethical implications of GenAI in mental health highlight the urgent need for further research at the intersection of human-AI interaction and psychotherapy professional practice. Currently, many commercial products and research prototypes rely on simple disclaimers (e.g., ``the bot may make mistakes'')~\cite{techtimes_sharma2024aibot} to convey limitations. However, we argue that such passive statements are rarely sufficient as they fail to offer actionable guidance or illustrate how users might navigate complex or ambiguous situations. 
Rather than relying solely on passive disclaimers, future research should explore actionable design strategies that ethically and responsibly mitigate the risks associated with integrating GenAI into mental health support. For example, perhaps, GenAI tools could incorporate interactive educational prompts or step-by-step guidance that encourages users to seek professional input; or to explore how GenAI agents can be designed to consult with human psychotherapists (and in what contexts), either directly or through referral mechanisms. Embedding psychoeducational content alongside AI-generated suggestions may also help users understand a chatbot's limitations, prompting them to verify critical information and maintain a balanced perspective. 

While static evaluation measures can provide an initial (actionable) framework for addressing ethical concerns, they may not fully capture the complexity introduced by GenAI's evolving nature. Unannounced updates or subtle algorithmic changes can alter a chatbot's tone or responsiveness, potentially undermining trust and increasing the risk of harm, especially for vulnerable individuals. To mitigate these issues, HCI researchers and practitioners should adopt user-centered, adaptive approaches that evolve alongside the technology. Longitudinal research will be crucial in continually assessing model impacts, informing refinements, and ensuring that GenAI remains a supportive, rather than destabilizing, complement to mental health care.

Additionally, on the policy and professional practice side, it is important to establish currently missing accreditation standards or professional guidelines for GenAI-driven tools in therapeutic contexts would help ensure adherence to rigorous ethical and clinical benchmarks. Regulatory frameworks, for example, might mandate transparency about model updates, require developers to implement safeguards that detect and correct detrimental shifts in empathetic or reliable responses, and reinforce data protection and privacy measures. Involving mental health professionals, ethicists, HCI researchers, AI developers, and user advocacy groups in participatory design and oversight committees would align GenAI systems with core therapeutic principles and uphold user trust and safety.

\subsubsection{The Evolving Client–Psychotherapist–GenAI Paradigm}
The growing integration of GenAI tools into mental health care signals a shifting dynamic in the client–psychotherapist relationship, a concept historically understood as a core component of effective therapy~\cite{bohart2013client, stein1984relationship}. Traditionally, the therapeutic alliance---characterized by trust, empathy, and mutual agreement on treatment goals---has been central to client outcomes, with extensive evidence linking a strong alliance to improved mental health results~\cite{arnow2013relationship}. In this evolving landscape, however, algorithmic interventions now interpose new layers of complexity, potentially recasting how clients and therapists perceive and engage with one another. This shift raises more pressing questions than definitive answers: How will GenAI influence the (subtle) interpersonal dynamics long considered foundational to therapy? Under what circumstances might these changes yield genuine benefits, and when could they prove detrimental? Most critically, how can human therapists oversee and evaluate GenAI's influence on patient well-being, ensuring that technological tools complement rather than erode the professional care and relational depth at the heart of effective psychotherapy? 

Answering these questions requires systematic empirical inquiry and thoughtful design exploration. Future studies might explore how clients perceive the boundaries between human expertise and AI-generated suggestions, how therapists assess the authenticity and appropriateness of chatbot responses, and how both parties negotiate trust and authority in the presence of AI. Through a series of inquiries, researchers can identify the circumstances under which GenAI supports or undermines existing therapeutic processes, and begin to develop measures, both qualitative and quantitative, that capture changes in therapeutic alliance, client engagement, and long-term well-being. In terms of potential design, HCI researchers could collaborate with therapists through participatory methods~\cite{spinuzzi2005methodology} to develop systems, for example, to support ``between-session'' psychoeducational skill-building homework—structured activities that clients complete independently but discuss in therapy sessions~\cite{tang2017supporting}. Such platforms could help therapists better understand GenAI's capabilities and limitations, teach them how to interpret AI-driven suggestions, and guide them in setting appropriate boundaries for client use.

%----------------------------------------------------------------------------------------------------
\section{Limitations}
A limitation of this study is the reliance on self-reported data, which may be subject to response biases such as social desirability or recall inaccuracies. Additionally, while our interviews provided in-depth insights, the relatively small and specific sample size may limit the generalizability of the findings. The study's focus on emotional trust and SA within a particular context (GenAI chatbot usage) may also not fully capture the complexities of these dynamics in broader or different contexts. For example, the absence of a group with high SA severity who have never used GenAI may limit the generalizability of the results across all symptom severity and usage duration combinations. Future research with more diverse participant pools is needed to ascertain whether such a subgroup would yield different engagement patterns. Furthermore, the scales used in our study require ecological validation to ensure their reliability and effectiveness in other real-world settings, such as diverse social contexts and different populations. The rapidly evolving nature of GenAI chatbot technology means that the findings may not fully account for future developments or changes in how such tools are used and perceived. Lastly, while we explored various dimensions of trust, other relevant factors, such as the influence of cultural differences on trust in AI, were not deeply examined and warrant further investigation.

%----------------------------------------------------------------------------------------------------
\section{Conclusion}
Through a mixed-methods study, our results show a strong correlation between trust in GenAI chatbots and willingness to use them for managing SA, with individuals showing more severe symptoms expressing a higher propensity to adopt GenAI chatbot solutions. Further, our qualitative findings revealed that trust dynamics and willingness to use GenAI chatbots varied with individuals’ severity of symptoms and their prior experiences with human psychotherapists and GenAI chatbots. Particularly, participants with severe symptoms appreciated GenAI chatbots' empathetic and non-judgmental support, emphasizing the importance of emotional trust. However, those with milder symptoms prioritized cognitive trust, focusing on GenAI chatbots' technical reliability and accuracy. These findings highlight the nuanced needs of users and suggest that future advancements in GenAI chatbots should focus on enhancing not only cognitive trust but also emotional trust to better support individuals in dealing with SA challenges.

\begin{acks}
We thank our anonymous reviewers for their reviews. This work is supported by the National Science Foundation for support under award no. NSF-2418582, the Accelerate Foundation Models Research (AFMR) of Microsoft Research, and the Society of 1918.  
\end{acks}


\bibliographystyle{ACM-Reference-Format}
\bibliography{ref}

\appendix
\section{Appendix}
\label{sec:appendices}

%TC:ignore
\subsection{Survey measures}
\textbf{Social anxiety:}

1) \textit{Have you ever been diagnosed with a social anxiety condition by a healthcare professional?} (0 = ``No,'' 1 = ``Yes.'')

2) \textit{Do you consider yourself to have social anxiety?} (0 = ``No,'' 1 = ``Yes,'' 2 = ``Not Sure.'')

3) Social Phobia Inventory (SPIN)~\cite{Connor_Davidson_Churchill_Sherwood_Weisler_Foa_2000} is a validated scale used to assess the severity of social anxiety symptoms. The SPIN consists of 17 items, each rated on a scale from 0 (not at all) to 4 (extremely), resulting in total scores ranging from 0 to 68. Scoring involves summing all the items, with a score above 20 suggesting the possibility of social anxiety. In research contexts, this threshold has been effective in distinguishing between individuals with social phobia and those in a control group. Items include a series of statements that respondents rate to reflect their experiences and feelings in social situations, including: 
\begin{enumerate}
    \item I am afraid of people in authority.
    \item I am bothered by blushing in front of people.
    \item Parties and social events scare me.
    \item I avoid talking to people I don’t know.
    \item Being criticized scares me a lot.
    \item Fear of embarrassment causes me to avoid doing things or speaking to people.
    \item Sweating in front of others causes me distress.
    \item I avoid going to parties.
    \item I avoid activities in which I am the center of attention.
    \item Talking to strangers scares me.
    \item I avoid having to give speeches.
    \item I would do anything to avoid being criticized.
    \item Heart palpitations bother me when I am around people.
    \item I am afraid of doing things when people might be watching.
    \item Being embarrassed or looking stupid are among my worst fears.
    \item I avoid speaking to anyone in authority.
    \item Trembling or shaking in front of others is distressing to me.
\end{enumerate}

\begin{table}[h]
\centering
\caption{SPIN Scoring and Symptom Severity}
\begin{tabular}{cl}
    \toprule
    \textbf{Score} & \textbf{Symptom Severity} \\
    \midrule
    0 - 20 & None \\
    21 - 30 & Mild \\
    31 - 40 & Moderate \\
    41 - 50 & Severe \\
    51 - 68 & Very severe \\
    \bottomrule
\end{tabular}
\end{table}

\textbf{Coping strategies:}

1) \textit{Have you sought external support or resources for managing your social anxiety?} ( Participants were allowed to select all applicable options from the following list: ``No,'' ``Therapy,'' ``Medication,'' ``GenAI Chatbots or digital mental health apps,'' ``Others.'')

2) \textit{Please select any of the following barriers you experience when seeking help for your mental health.} ( Participants were allowed to select all applicable options from the following list ``Cultural limitations,'' ``Social or family pressures,'' ``Financial constraints,'' ``Feelings of shame,'' ``Fear of stigma,'' ``Lack of time,'' ``Language barriers,'' ``Physical health concerns,'' ``I do not experience any barriers,'' ``Others.'')

\textbf{Past experiences with GenAI chatbots:}

1) \textit{How long have you been using GenAI chatbots?} (Responses were categorized into time frames such as ``Never,'' ``Less than 3 month,'' ``3-6 months,'' ``6-12 months,'' ``More than one year.'')

2) \textit{In the past 6 months, how often have you used a GenAI chatbot for social anxiety support?} (Participants could choose from options such as ``Never,''``Rarely (once or twice a month),'' ``Occasionally (a few times a month),'' ``Frequently (once or twice a week),'' ``Very frequently (several times a week).'')

\textbf{Willingness to use GenAI chatbots for SA support:}

\textit{Would you be open to using a specially designed GenAI chatbot as a supportive tool for managing social anxiety?} (0 = ``No, I would not be interested,'' 1 = ``Yes,'' 2 = ``Possibly, I am open to use.'')

\textbf{Trust in GenAI chatbots for social anxiety support}

1) \textit{I believe GenAI chatbots can provide accurate and helpful information.} 

\textit{Rationale:} This measure reflects the dimension of competence, which pertains to the users' perceptions of the GenAI chatbots' ability to provide accurate, reliable, and useful information and confidence that the GenAI chatbot is knowledgeable and capable of assisting them effectively in managing their social anxiety.

2) \textit{I trust GenAI chatbots to handle my sensitive or personal information securely.} 

\textit{Rationale:} This measure captures the dimension of honesty, which relates to users' beliefs about the chatbot's integrity and transparency in handling their personal data. Trust in this context is built on the assurance that the chatbot will protect their privacy and not misuse or disclose their sensitive information without consent.

3) \textit{I would be willing to follow a suggestion or recommendation made by a GenAI chatbot, even if I was unsure if it was the best choice.} 

\textit{Rationale:} This measure reflects the dimension of experience, which relates to users' past interactions with the chatbot and how these experiences influence their willingness to trust its recommendations.

4) \textit{I believe GenAI chatbots operate in my best interest.} 

\textit{Rationale:} This measure captures the dimension of benevolence, which pertains to users' perceptions of the chatbot's intentions and the belief that the chatbot is designed to support and prioritize the user's mental health, acting in their best interest.

5) \textit{I feel that GenAI chatbots understand my needs and respond appropriately.} 

\textit{Rationale:} This measure reflects the dimension of reliability, which involves the users' expectation that the chatbot will consistently respond in a manner that is attuned to the user's specific circumstances and requirements.

6) \textit{I think GenAI chatbots will become more trustworthy as technology advances and they become more sophisticated.} 

\textit{Rationale:} This measure captures the dimension of expectation, which relates to users' forward-looking beliefs about the future capabilities of GenAI chatbots. 

\textbf{Demographic background}

1) \textit{Age.} We collected participants' ages as numeric data, which were categorized into two groups (18-24 and 25+).

2) \textit{Gender.} Participants were given the options of ``Woman,'' ``Man,'' ``Non-binary,'' and ``Prefer not to answer.''

3) \textit{Education.} Participants were asked about their highest level of education using the question, \textit{``What is the highest degree or level of school you have completed?''} The response options included: ``High school graduate,'' ``Associate degree,'' ``Bachelor’s degree,'' ``Master’s degree,'' ``Doctorate or professional degree,'' or ``Prefer not to answer.''

4) \textit{Race.} Participants could select from the options for race: ``White,'' ``African American or Black,'' ``Asian,'' ``Other,'' and ``Prefer not to answer.''

\subsection{Supplement for Survey Results}

\begin{figure*}[h!]
\centering
\includegraphics[width=1\linewidth]{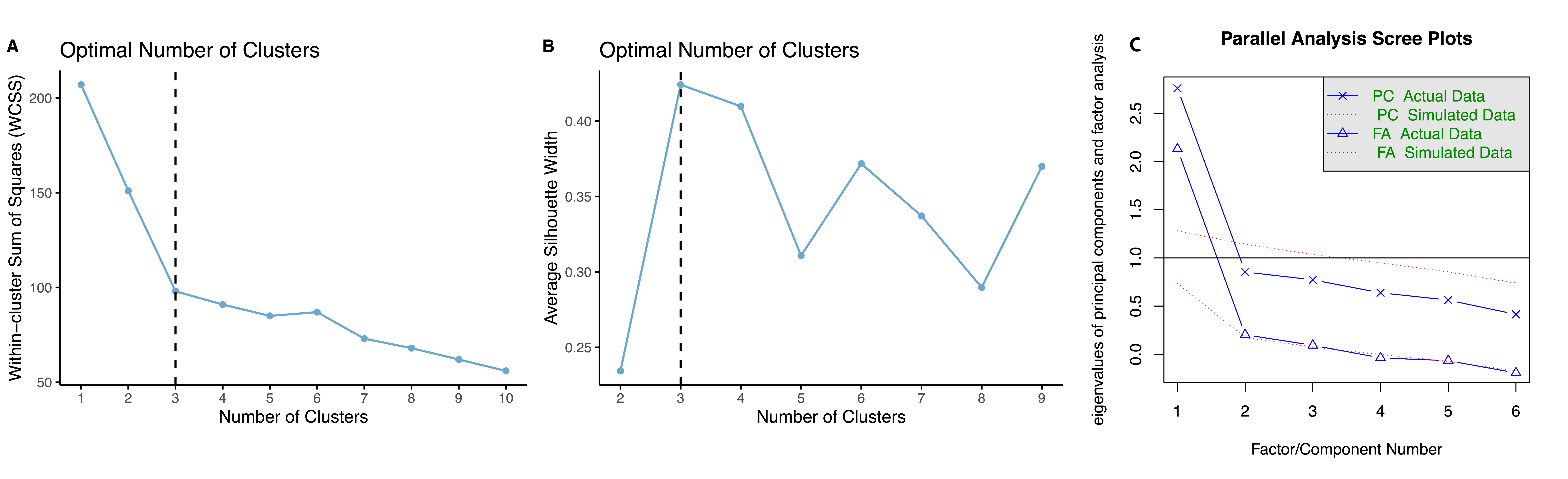}
\caption{Cluster and component definition criteria: (A) Within-Sum-of-Squares (WSS) for GMM clustering, (B) Average Silhouette Method for GMM clustering, and (C) Parallel analysis for determining the number of components to retain in factor analysis.}
\label{fig:criteria}
\end{figure*}

\begin{figure*}[h!]
\centering
\includegraphics[width=0.9\linewidth]{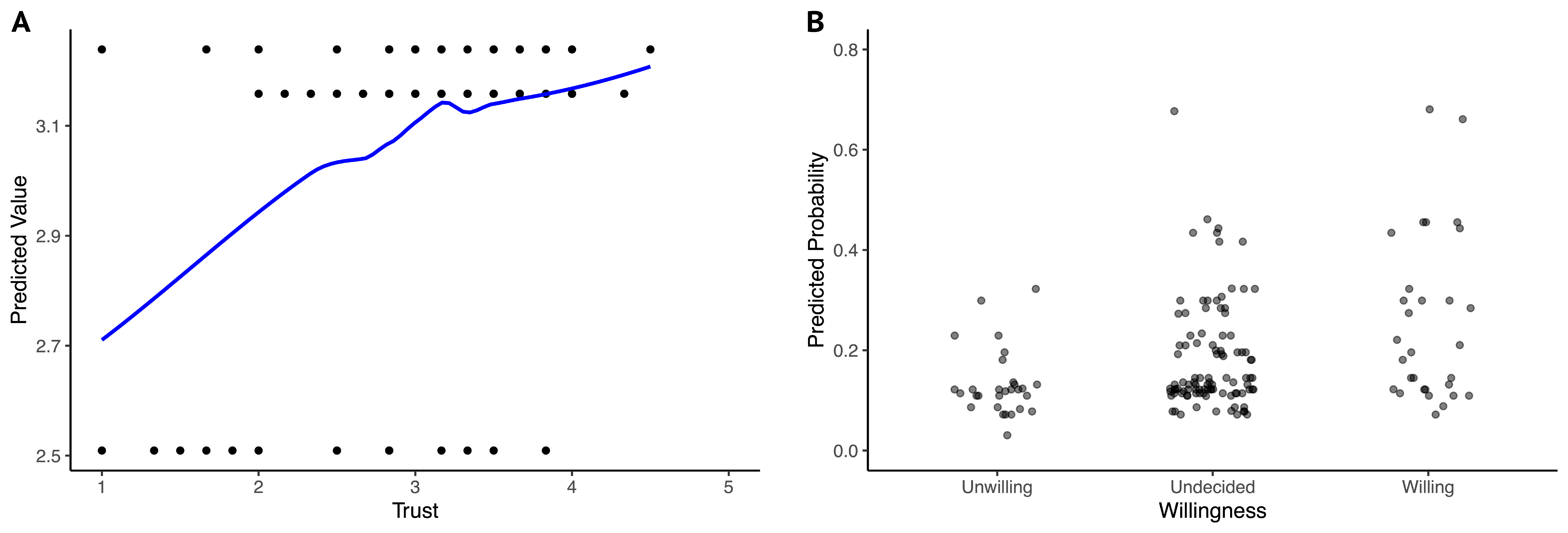}
\caption{Scatter plot (A) illustrating the relationship between predicted values and actual Trust, based on a linear regression model of Trust and Willingness; (B) illustrating the relationship between predicted willingness to use and actual willingness, based on an ordinal logistic regression model explaining respondents’ willingness to use GenAI chatbots for SA support.}
\label{fig:visual}
\end{figure*}

%TC:endignore

\end{document}